\documentclass[superscriptaddress, amsmath,amssymb]{revtex4-1}
\usepackage{graphicx}
\usepackage{amsmath, amsthm, amssymb}
\usepackage{latexsym}
\usepackage{amssymb}
\usepackage{bm}
\usepackage{dcolumn}
\usepackage{epstopdf}
\usepackage{color}
\usepackage{braket}
\usepackage[normalem]{ulem}
\usepackage{float}

\newcommand{\YPP}{Yb$_2$Pt$_2$Pb}
\newcommand{\ql}{$\mathbf{q}_{L}$}
\newcommand{\qh}{$\mathbf{q}_{HH}$}
\newcommand{\kf}{$\mathbf{k}_\mathrm{F}$}
\newcommand{\bq}{\mbox{$\mathbf{q}$}}
\newcommand{\bS}{\mbox{$\mathbf{S}$}}
\newcommand{\bH}{\mbox{$\mathbf{H}$}}
\newcommand{\cmmnt}[1]{}

\begin{document}

\title{Spinon Confinement and a Sharp Longitudinal Mode in Yb$_2$Pt$_2$Pb in Magnetic Fields}
\date{\today}

\author{W. J. Gannon}
\altaffiliation[Present address: ]{Stewart Blusson Quantum Matter Institute, University of British Columbia, Vancouver, British Columbia V6T 1Z4, Canada}

\affiliation{Department of Physics and Astronomy, Texas A\&M University, College Station, Texas 77843, USA}

\author{I. A. Zaliznyak}
\affiliation{Condensed Matter Physics and Materials Science Division, Brookhaven National Laboratory, Upton, New York 11973, USA }

\author{L. S. Wu}
\altaffiliation[Present address: ]{Department of Physics, South University of Science and Technology of China, Shenzhen,
518055, China}
\affiliation{Quantum Condensed Matter Division, Oak Ridge National Laboratory, Oak Ridge, Tennessee 37830, USA}

\author{A. E. Feiguin}
\affiliation{Department of Physics, Northeastern University, Boston, Massachusetts 02115, USA}

\author{A. M. Tsvelik}
\affiliation{Condensed Matter Physics and Materials Science Division, Brookhaven National Laboratory, Upton, New York 11973, USA }

\author{F. Demmel}
\affiliation{ISIS Facility, Rutherford Appleton Laboratory, Didcot, OX11 0QZ, United Kingdom}

\author{Y. Qiu}
\affiliation{NIST Center for Neutron Research, National Institute of Standards and Technology, Gaithersburg, MD 20899, USA}

\author{J. R. D. Copley}
\affiliation{NIST Center for Neutron Research, National Institute of Standards and Technology, Gaithersburg, MD 20899, USA}

\author{M. S. Kim}
\affiliation{Department of Physics and Astronomy, Stony Brook University, Stony Brook, New York 11794, USA}

\author{M. C. Aronson}
\altaffiliation[Present address: ]{Stewart Blusson Quantum Matter Institute, University of British Columbia, Vancouver, British Columbia V6T 1Z4, Canada}
\affiliation{Department of Physics and Astronomy, Texas A\&M University, College Station, Texas 77843, USA}

\begin{abstract}

The fundamental excitations in an antiferromagnetic chain of spins-1/2 are spinons, de-confined fractional quasiparticles that when combined in pairs, form a triplet excitation continuum. In an Ising-like spin chain the continuum is gapped and the ground state is N{\'e}el ordered. Here, we report high resolution neutron scattering experiments, which reveal how a magnetic field closes this gap and drives the spin chains in \YPP\ to a critical, disordered Luttinger-liquid state. In \YPP\ the effective spins-1/2 describe the dynamics of large, Ising-like Yb magnetic moments, ensuring that the measured excitations are exclusively longitudinal, which we find to be well described by time-dependent density matrix renormalization group calculations. The inter-chain coupling leads to the confinement of spinons, a condensed matter analog of quark confinement in quantum chromodynamics. Insensitive to transverse fluctuations, our measurements show how a gapless, dispersive longitudinal mode arises from confinement and evolves with magnetic order.

\end{abstract}

\maketitle

\section*{Introduction}
The one-dimensional (1D) \textit{XXZ} Hamiltonian for quantum spin chains given by Eqn.~\ref{XXZ} is a cradle of exactly solvable quantum theory models of interacting many-body systems~\cite{Bethe_Zphys_1931}. The exact solution features purely quantum-mechanical entities and concepts such as fractional excitations and the quantum-critical Luttinger-liquid state \cite{Haldane_PRL1980,Fadeev_PhysLett1981,Polyakov_NucPhys1977,Shelton_PRB_1996,Lake_NatMat_2005,Zaliznyak_NatMat2005,Lake_NatPhys_2009,Tsvelik_book_2003}. The Hamiltonian considers the components $S_i^{\alpha}$ ($\alpha={x,y,z}$) of a spin angular momentum operator, $\bS_i$ ($S=1/2$) at site $i$ on a 1D chain, with $\mathcal{J}$ a nearest neighbor exchange coupling for $x,y$ spin components, $\Delta$ a uniaxial coupling anisotropy, and $\bH$ magnetic field (with $g$ and $\mu_B$ the Lande g-factor and Bohr magneton respectively),
\begin{equation}\label{XXZ}
\mathcal{H}=\mathcal{J}\sum_{i} \left( S^{x}_{i}S^{x}_{i+1} + S^{y}_{i}S^{y}_{i+1} \right) + \Delta S^{z}_{i}S^{z}_{i+1}- g\mu_B \sum_{i}\bH\cdot \bS_i.
\end{equation}
The low energy excitations of this model (\ref{XXZ}) are spin-1/2 quasiparticles called spinons. In the limit of strong Ising anisotropy, $\Delta \gg 1$, spinons can be visualized as domain walls in an antiferromagnetically ordered ground state of the chain [Fig.~\ref{ZeroField}(A)]. Angular momentum conservation mandates that spinons are always created in pairs, such that each spinon carries a fraction, $\pm 1/2$, of the angular momentum change, $\Delta S^z = 0, \pm 1$, required to initially introduce the domain walls in an infinite chain. Since moving these domain walls is an energy and angular momentum conserving process, the walls will propagate freely, carrying the quanta of energy, $E$, and linear momentum, $q$, introduced by their creation [Fig~\ref{ZeroField}(A)]. The physics contained in Eqn.~\ref{XXZ} leads directly to the separation of the spin from other electronic degrees of freedom, mapping directly onto that of the Luttinger liquid for $-1 \leq \Delta \leq 1$~\cite{Haldane_PRL1980,Lake_NatMat_2005,Zaliznyak_NatMat2005,Lake_NatPhys_2009,Tsvelik_book_2003,Giamarchi_book_2004}.

Coupling the chains described by Eqn.~\ref{XXZ} leads to new and emergent physics. Analogous to quark confinement in quantum chromodynamics\cite{Polyakov_NucPhys1977,Shelton_PRB_1996,Lake_NatPhys_2009}, the dimensional crossover from 1D chains to 3D coupled chains leads to quasiparticle confinement, thereby stabilizing long range magnetic order at temperature $T>0$. A new excitation of the longitudinal degree of freedom of the order parameter is predicted when the interchain coupling is weak \cite{Lake_NatMat_2005}.  These phenomena have been the subject of a considerable amount amount of recent experimental work in \textit{XXZ} spin chain materials~\cite{Grenier_PRL_2015, Matsuda_PRB_2017, Faure_NatPhys_2018, Wang_PRB_2015, Bera_PRB_2017}. Like a similar longitudinal mode previously observed near the critical point in a system of coupled spin-1/2 dimers ~\cite{Ruegg_PRL_2008, Merchant_NatPhys_2014}, this excitation can be interpreted as a condensed matter analog of the Higgs boson~\cite{Pekker_ARCMP_2015}. The neutron scattering experiments reported here on the one dimensional rare-earth antiferromagnet \YPP\ investigate these fundamental processes in detail, using an external magnetic field as a tuning parameter.

\section*{Results}
\subsection*{Inelastic Neutron Scattering on \YPP}
\begin{figure}
\includegraphics[height=100mm]{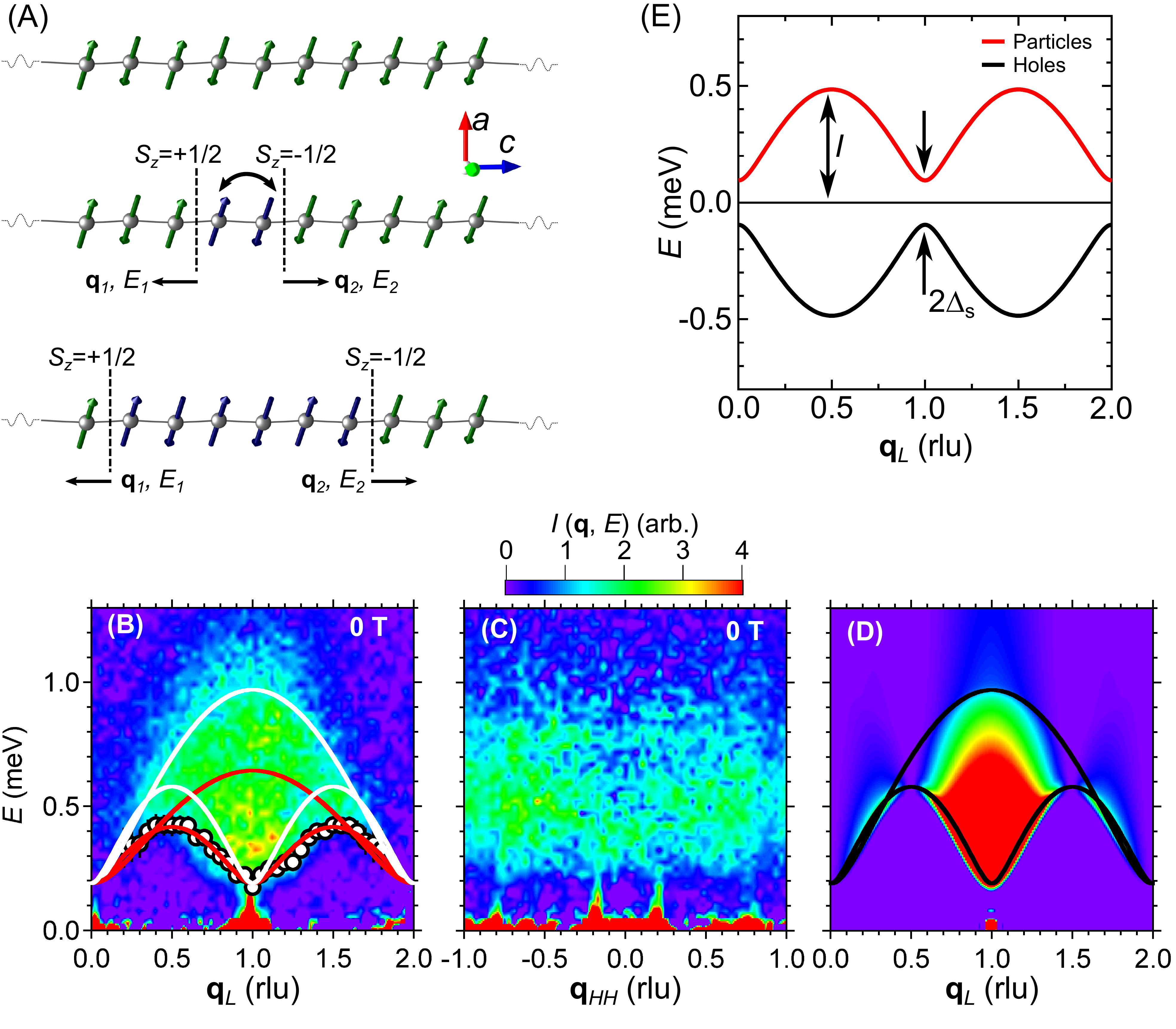}[H]
\caption{\label{ZeroField}\textbf{Spinons on decoupled chains} (A) An anisotropic AFM spin chain (top). If two spins are interchanged, two domain walls are created between the original antiferromagnetic domain (green) and a new domain (blue) (middle). Those domain walls form the basis for the propagating states carrying energy and momentum quanta (bottom). (B) The magnetic excitation spectrum and dispersion along the \ql\ direction in reciprocal space measured at $T = 0.1$~K, summed over $-1\leq \mathbf{q}_{HH} \leq1$ rlu. The lower boundary of the spectrum (white circles) is shown along with the boundaries of the two spinon continuum obtained by fitting the lower boundary (red lines, $\Delta=3.46$) and comparing the total measured spectrum to theory (white lines, $\Delta=2.6$) \cite{Wu_Science_2016}. The color scale for parts B-D is shown above part C.  Error bars represent one standard deviation.  (C) The magnetic excitation spectrum along the \qh\ direction in reciprocal space measured at $T=0.1~\rm K$, summed over $0\leq 	\mathbf{q}_L \leq2$ r.l.u. (D) The spinon spectrum obtained from tDMRG calculations for the {\it XXZ} model (\ref{XXZ}) with $\Delta=2.6$ on the 96-site chain. The continuum boundary (black lines) is the same as that shown in (B) for $\Delta=2.6$.   (E) The dispersions of particle-like (red) and hole-like (black) spinons, symmetric about $E=0$ in zero magnetic field, are sketched with the real  $\Delta=2.6$ parameters for \YPP. The bandwidth parameter $I$ and the spinon gap $\Delta_\mathrm{S}$ are indicated by arrows, with $2\Delta_\mathrm{S}$ the energy separation between the particle and hole bands at $\mathbf{q}_L$=0, 1, and 2 rlu.}
\end{figure}

\YPP\ is a metal with a planar crystal structure where orthogonal pairs of Yb ions form a Shastry-Sutherland lattice (SSL) motif in the tetragonal $\mathbf{a}-\mathbf{b}$ plane \cite{Shastry_PB_1981, Pottgen_JSSC_1999, Kim_PRB_2008, Ochiai_JPSJ_2011, Shimura_JPSJ_2012, Iwakawa_JPSJ_2012, Kim_PRL_2013}. High resolution neutron scattering experiments recently showed that the physics of $4f$-orbital overlaps leads to  unusual consequences for the magnetism in \YPP~ \cite{Wu_Science_2016}.  The low energy magnetic excitations are spinons, having a quantum continuum for momentum along the chain direction \ql\ that can be measured with inelastic neutron scattering,~\cite{Squires_book_1978} with an excitation bandwidth that is considerably larger than the excitation gap [Fig.~\ref{ZeroField}(B)].  For momenta in the interchain \qh\ direction [Fig.~\ref{ZeroField}(C)], the continuum is entirely flat,  indicating that the spinons are completely incoherent between the chains.

In zero field, our measurements agree well with time-dependent density matrix renormalization group (tDMRG) calculations~\cite{White2004} for the {\it XXZ} model (\ref{XXZ}) [Fig.~\ref{ZeroField}(D)], although experiment indicates that the spectral weight is spread throughout the spinon Brillouin zone (BZ) more evenly and to higher energies than these calculations predict, suggesting non-negligible next-neighbor coupling \cite{Wu_Science_2016}. Comparisons of our data to theory indicate only a modest anisotropy, $\Delta \sim 2-3$. It is clear that the $XXZ$ Hamiltonian (\ref{XXZ}) is an appropriate description for \YPP\ despite the large and orbitally dominated moment of the Yb ions. Due to their Kramers doublet ground state of almost pure $\ket{J, m_J}=\ket{7/2, \pm 7/2}$, the Yb moments have a pseudospin $\mathbf{S}= 1/2$ character \cite{Wu_Science_2016,Miiller_PRB_2016}. Rather than quenching the quantum spin dynamics, the strong Ising magnetic anisotropy imposed by the crystal electric field acting on the $f-$orbital wave function instead singles out the longitudinal excitation channel in two orthogonal sublattices of 1D chains with moments oriented along the (110) and ({\=1}10) crystal directions.

The essential features of the quantum continuum can be understood by noting that each spinon carries spin $1/2$, and so the angular momentum selection rules dictate that neutron scattering in \YPP\ measures two-spinon states where the total spin is zero, with one spinon in each spin state, $\pm 1/2$.  In order to describe the boundaries of the two-spinon continuum, it is convenient to adopt the language of particles and holes occupying the fermionic spinon dispersion along the chain direction, $E_{\mathrm{p,h}} = \pm \left(I^2\sin^2\left(\pi \mathbf{q}_L\right)+\Delta_{\mathrm{S}}^2\cos^2\left(\pi \mathbf{q}_L\right)\right)^{1/2}$, $0 \leq \mathbf{q}_L < 1~\rm{rlu}$, where $\Delta_\mathrm{S}$ is an energy gap brought on by the {\it XXZ} anisotropy $\Delta > 1$, and $I$ defines the dispersion bandwidth and encodes the coupling $\mathcal{J}$ [Fig.~\ref{ZeroField}(E)]~\cite{Bethe_Zphys_1931, Wu_Science_2016, Bougourzi_PRB1998, Caux_JStatMech_2006}. In place of electric charge, these particles and holes each carry a half unit of spin angular momentum.  The boundaries of the two-spinon continuum are defined by the extremal energy and momentum conserving combinations of one particle and one hole, and they are shown in Fig.~\ref{ZeroField}(B) for both $\Delta=2.6$ and 3.46, the range of values determined in previous work~\cite{Wu_Science_2016} (see Supplementary Note 1).  At zero magnetic field, the chemical potential is in the middle of the gap separating the particle and the hole bands, which describes the antiferromagnetic (AFM) state with zero total spin, $S^z = 0$.  The size of the $T=0$ ordered moment implied by the $XXZ$ anisotropy is consistent with our measurements at $T=0.1$ K,  within the precision of our data~\cite{Wu_Science_2016}.  This implies that the interchain coupling responsible for moving the N\'eel temperature away from $T=0$, the value predicted by the $XXZ$ model, to $T_\mathrm{N}=2.07$ K is less than both the intrachain exchange $\mathcal{J} = 0.206~\mathrm{meV}$ and the spinon gap $\Delta_\mathrm{S}=0.095~{\rm meV}$, the dominant 1D energy scales. The flat dispersion of the excitations between the chains in zero field [Fig.~\ref{ZeroField}(C)], despite the apparent ladder geometry of the crystal structure, suggests that the effect of interchain interactions on low energy excitations is quenched when $\Delta_\mathrm{S}$ is nonzero.

\begin{figure}
\includegraphics[height=140mm]{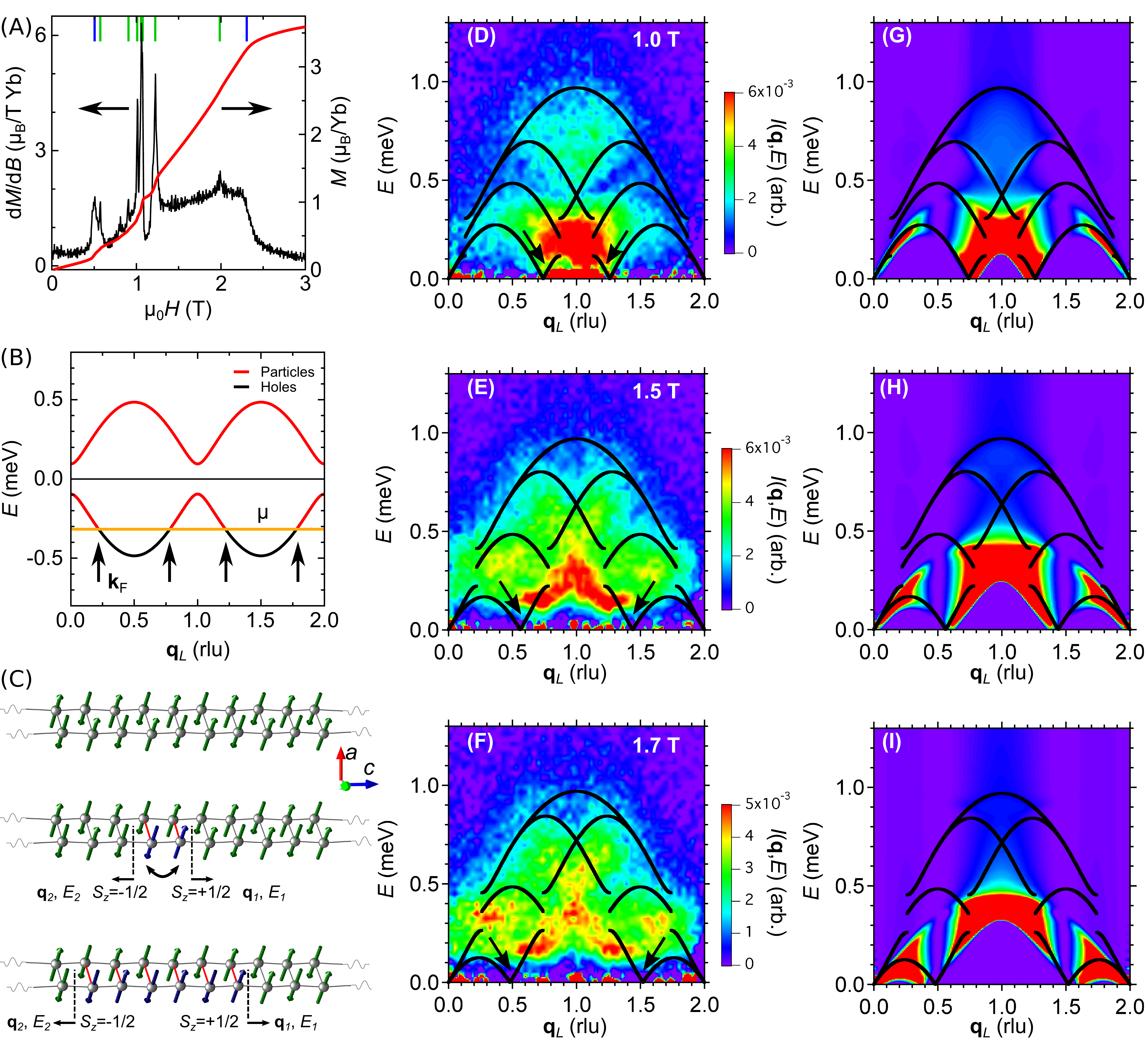}[H]
\caption{\label{InField1}\textbf{Spinons in a magnetic field}. (A) The magnetic field dependence of the static magnetization at $k_{\mathrm{B}}T<\Delta_\mathrm{S}$ [red, right axis, $\mathbf{H}||(110)$] shows several discontinuous jumps (blue and green dashes), corresponding to transitions among different 3D ordered phases that are more clear in $\mathrm{d}M/\mathrm{d}\mu_0H$ (black, left axis) \cite{Shimura_JPSJ_2012, Kim_PRL_2013}. These phases correspond to different ways that magnetic moments arrange into registry minimizing the energy of magnetic dipole interactions between the Yb moments. (B) When a magnetic field is applied along the chain direction, the chemical potential $\mu = -g\mu_BHS^z$ (yellow) is lowered, emptying part of the hole band when $\left | \mu \right | > \left |\Delta_\mathrm{S} \right |$. $\mu$ crosses the hole dispersion at four points in the Brillouin zone (black arrows), defining the Fermi wavevector \kf.  (C) Two AFM ordered, 1D spin chains (top). If two spins on one chain are interchanged, two domain walls are formed between the original domain (green) and a new one (blue). The new domain frustrates the interchain interaction, represented by the red interchain bonds.  This frustration creates a linear potential confining low energy spinons to bound states (bottom). (D-F) The magnetic excitation spectrum and its dispersion along the \ql\ direction in reciprocal space measured at $T = 0.1~\rm K$ and  $\mu_0 H=1.0~\rm T$ (D), 1.5 T (E), and 1.7 T (F), summed over $-1\leq \mathbf{q}_{HH} \leq1$ rlu.  The dispersions for the extremal combinations of particles and holes are shown (black lines) (See Supplementary Note 2). The spinon bound states are manifest from the enhanced low energy spectral weight around $1 \pm 2$\kf\ (black arrows). (G-I) The spinon spectrum computed using tDMRG calculations for the {\it XXZ} model (\ref{XXZ}) on a 96-site chain with $\Delta=2.6$ at equivalent chain magnetizations as (D-F), shown on the same color scale. The dispersions for the extremal combinations of particles and holes are also shown as black lines. }
\end{figure}

A magnetic field along the $\mathbf{z}$ (110) direction introduces the Zeeman term $-g \mu_B H \sum_{i}S^{z}_i$ to Eqn.~(\ref{XXZ}), which lowers the chemical potential, $\mu = -g\mu_BHS^z$. The potential needed to close the energy gap for creating a hole on the spinon dispersion is $\left|\mu \right| =\Delta_\mathrm{S}=0.095~\mathrm{meV}$.  Taking $g=7.3$~\cite{Wu_Science_2016}, $\left|\mu \right|=\Delta_\mathrm{S}$ corresponds to a critical field of $\mu_0 H = 0.5~\rm T$.  An abrupt increase in the bulk magnetization is seen at this field when oriented parallel to the magnetic moments of either sublattice at temperatures $k_BT<\Delta_\mathrm{S}$ [Fig.~\ref{InField1}(A)]~\cite{Shimura_JPSJ_2012, Kim_PRL_2013}. On the other extreme of the magnetization, when $\left|\mu \right| > I$, the entire hole band lies above the chemical potential, the field having transformed all holes to particles. Particle-hole pairs can no longer be produced, quenching spinon excitations, and producing a ferromagnetic (FM) state. The saturation field in \YPP\ is 2.3 T, precisely the field needed for $\mu=0.485~\mathrm{meV}$, the value of $I$ when $\Delta=2.6$, the number obtained by requiring that Eqn.~(\ref{XXZ}) provides best description of the entire $\mu_0 H = 0$ excitation spectrum~\cite{Wu_Science_2016}. The comparison is less favorable when $\Delta$ is taken to be 3.46, the value derived directly from fitting the lower boundary of the continuum.  As the static properties correspond to an integration over all energies, it is not surprising that they are better captured by $\Delta=2.6$ and we adopt this value of $\Delta$ here.

At intermediate fields, $0.5<\mu_0 H<2.3$ T, the hole band is partially emptied. The chemical potential crosses the hole dispersion at four points in the spinon BZ [Fig.~\ref{InField1}(B)],  defining a Fermi wavevector \kf\ that directly links particle and hole states. There are now eleven unique extremal states made from a single particle and hole, rather than the three that are possible in zero field. The boundaries of the two spinon continuum change dramatically, as the possible extremal states are heavily influenced by the restrictions of the hole energies and momenta and the additional phase space occupied by particles.

That is precisely what is measured in the neutron scattering spectra of \YPP\ [Fig.~\ref{InField1}(D-F)]. At $\mu_0 H=1.0$~T, there is strong scattering concentrated at low energies within a range $1 \pm 2$\kf\ around the BZ center, with a weaker continuum at higher energies. As the magnetic field is increased, the low-energy spectral weight spreads throughout the zone as \kf\ increases, with higher energy pockets of spectral weight bounded by the extremal two spinon states comprising the continuum. While the measured continua are in broad agreement with the results of tDMRG calculations performed on isolated chains [Fig.~\ref{InField1}(G-I)], there are marked differences at low energies where interchain interactions are important. The measured lower boundaries are gapped at small energies and also slightly distorted relative to theoretical expectations, with the increased spectral weight indicative of a bound state. This is a direct demonstration of spinon confinement induced by interchain coupling, which is not accounted for in the 1D calculations of Fig.~\ref{InField1}(G-I), and also shows how a magnetic field tunes the coexistence of confined and free spinons in \YPP\ for $\mu_0 H>0.5~\rm T$.

The schematic picture of spinons and their propagation presented in Fig.~\ref{ZeroField}(A) needs to be modified in the presence of the interchain interactions, where the creation of spinons on one chain leads to frustration of the AFM interactions between the chains. As two spinons separate, the energy of this frustration grows with the number of FM aligned neighbors [Fig.~\ref{InField1}(C)]. This provides a linear confining potential, just as quarks are confined by the gluon-mediated strong force in QCD, which also increases with quark separation.  When spinons are created with energies above the highest energy level existing in the confining potential introduced by the interchain coupling, these high energy quasiparticles  propagate freely within the two spinon continuum, demonstrating the same asymptotic freedom as experienced by unbound quarks~\cite{Greiner_book_1994}. A spinon bound state is observed in a neutron scattering experiment as excess spectral weight of resolution-limited energy width, which prominently appears near $1 \pm 2$\kf, the two soft spots around the BZ center, below the quantum continuum \cite{Giamarchi_book_2004}.

Perhaps the most interesting aspect of these free and confined excitations is how their dispersions develop in the \qh\ direction, perpendicular to the chains. Spinons are continually created in pairs on all chains, and in the absence of interchain coupling they are free to propagate. Order and frustration in coupled chains naturally lead to the bound states, which develop an interchain dispersion that reflects the underlying AFM order. Spinons are generated in registry on adjacent chains, thus minimizing the interchain frustration.

\begin{figure}
\includegraphics[height=140mm]{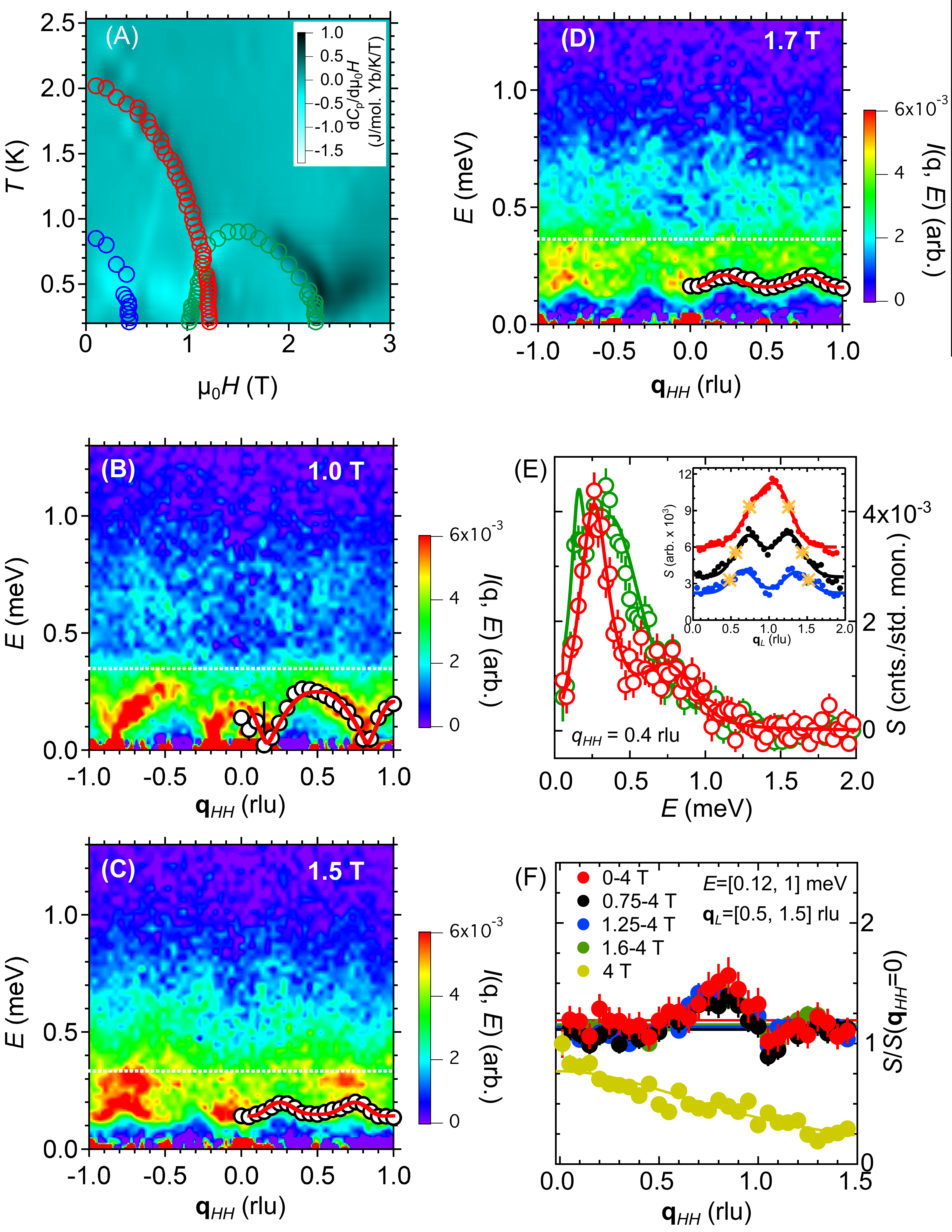}[H]
\caption{\label{InField2}\textbf{ A coherent, longitudinal interchain mode in \YPP.} (A) The field-temperature phase diagram of \YPP\ deduced from specific heat measurements (see Supplementary Note 3). Symbols representing phase lines of the low field AFM order (red), gapped phase when $k_{\mathrm{B}} T < \Delta_\mathrm{S}$ (blue), and second, weaker order (green) are obtained from the magnetic field dependence of the magnetization along the (110) direction at fixed temperatures.  Refer to Fig.~\ref{InField1}(A) for an example. Not all of the field-induced AFM phases observed in the  magnetization are shown. (B-D) The magnetic excitation spectrum along the \qh\ direction of reciprocal space measured at $T=0.1$~K and $\mu_0 H=1.0$~T (B), 1.5 T (C), and 1.7 T (D), summed over $0 \leq \mathbf{q}_L \leq 2$ rlu. The fitted dispersion of the interchain mode is shown (white circles), as well as fits to the expression described in the main text (red lines). The mode dispersion and fits are only shown for $\mathbf{q}_{HH}>0$ for clarity (see Supplementary Note 4).  The continuum boundary appears at $2\Delta_\mathrm{M}$~\cite{Schulz_PRL_1996} (white dashed line). (E) Example cuts of the energy dependence at $\mathbf{q}_{HH} = 0.4 ~\mathrm{rlu}$ measured at $B = 1.0$ (red) and 1.5 T (green). Fits are described in the text. The mode dispersions shown in (B-D) are  extracted from such fits. (Inset) The integrated intensity of the scattering over the energy of the interchain mode extracted from the fits is shown in parts B-D as a function of  momentum \ql\ along the chain.  Magnetic fields of 1.0 T (red), 1.5 T (black), and 1.7 T (blue) are offset for clarity. $\mathbf{q}_L = 1 \pm 2$\kf\ is shown at each field (gold stars). (F) The polarization of the excitations follows the projection onto the scattering vector $\mathbf{q}$ of the magnetic moments contributing to the scattering, indicating that all observed excitations are longitudinal (see Supplementary Note 5.) Fields and integration ranges are given in the figure legend; the calculated polarization was averaged within the corresponding \ql\ integration range. Error bars represent one standard deviation and where not visible are smaller than symbol size.}
\end{figure}

Fig.~\ref{InField2}(A) shows a complex phase diagram consisting of several distinct phases that differ markedly in the behavior of the inter-chain dispersion. Most notably, for $0.5 \lesssim \mu_0 H \lesssim 2.3$ T, we observe a new excitation that emerges from the featureless spinon continuum found along \qh\ in the gapped zero field N{\'e}el phase where the chains are effectively decoupled. This mode resides within the low-energy window of the two spinon bound states, but has a pronounced dispersion in the \qh\ (interchain) direction that changes considerably with increasing field [Fig.~\ref{InField2}(B-D)]. Remarkably, at 1.0 T the dispersive interchain mode appears nearly gapless, while its intensity is markedly larger than that of the continuum, Figs.~\ref{InField1}(D) and~\ref{InField2}(B). The mode becomes clearly gapped with increasing fields, while the relative spectral weight of the continuum grows.  We model the energy dependence of the scattering at a specific \qh\ as a damped harmonic oscillator (DHO) response centered at the mode position and the product of a Lorentzian and step function accounting for the continuum at higher energies, all convolved with the instrument resolution [Fig.~\ref{InField2}(E)].  The energy width of the new mode is roughly resolution limited at all fields, and it is always distinguishable from the spinon continuum for fields $\mu_0H\leq 1.7~\rm T$.  The connection of the mode and the confined spinon states can be emphasized by integrating over the energies of the mode and plotting the intensity as a function of momentum along the chains [Fig.~\ref{InField2}(E-Inset)]. At all fields, $\approx$85\% of the spectral weight is concentrated within the momentum range \ql$ = 1\pm 2$\kf.

Importantly, this interchain mode is longitudinally polarized. We confirm its longitudinal character by using the fact that neutron scattering cross-section is uniquely sensitive to magnetic fluctuations that are perpendicular to the wave vector transfer \cite{Squires_book_1978}. The intensity measured at 4T, in the FM state precisely follows the projection of the scattering wave vector on the (110) direction, revealing fluctuations polarized along the in-plane Ising moments, which are insensitive to magnetic fields [Fig.~\ref{InField2}(F)]. When this field-independent contribution is subtracted as a background, the resulting field-dependent intensity [Figs.~\ref{InField1} and \ref{InField2}] does not depend on the wave vector orientation in the scattering plane, indicating magnetic fluctuations polarized along the vertical direction, collinear with the magnetic field \cite{Wu_Science_2016} (see Supplementary Note 5).  At all fields, our measurements unambiguously probe the longitudinal response. The Ising anisotropy of the  $\ket{7/2, \pm 7/2}$ ground state doublet of the Yb moments nearly completely suppresses any transverse  magnetic fluctuations from our measurements.

\begin{figure}
\includegraphics[height=100mm]{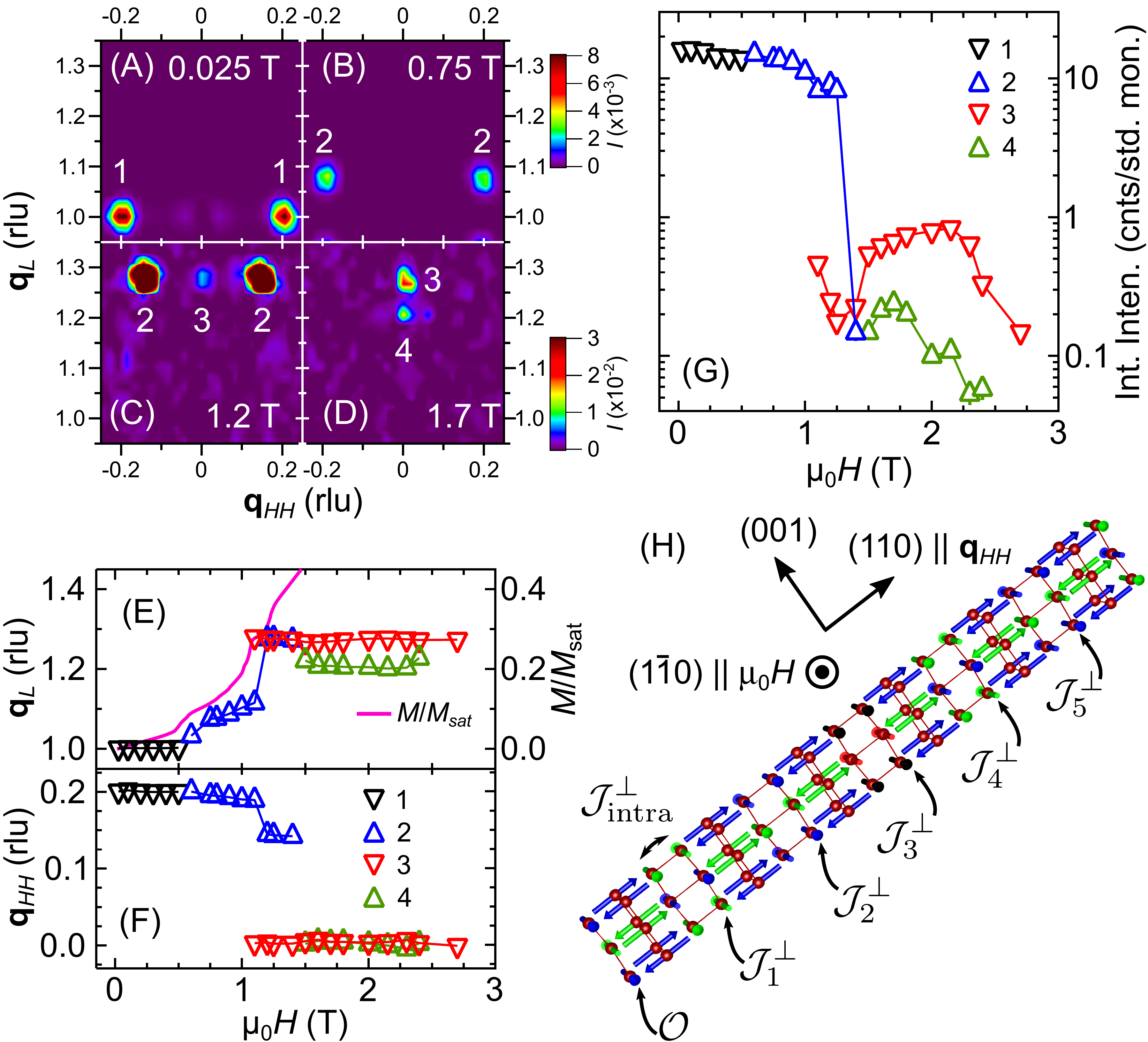}[H]
\caption{\label{Diffraction} \textbf{Magnetic order and phase diagram of \YPP}. (A-D). Time of flight data with energy transfer $E = 0$ in the \qh,\ql plane, measured at $T=0.1$ K in fields of 0.025 T (A), 0.75 T (B), 1.2 T (C), and 1.7 T (D). The magnetic Bragg diffraction peaks in each part of the phase diagram are labeled 1-4. (E-F) The location in reciprocal space along \ql\ (E) and \qh\ (F) of the magnetic Bragg scattering as a function of the field in the regions plotted in parts A-D. Data shown are an average, symmetrized about $\mathbf{q}_L=1$ and $\mathbf{q}_{HH}=0$. The open symbols correspond to the peaks labeled 1-4 in panels B-E, with black indicating peak 1, blue peak 2, red peak 3, and green peak 4. Magnetization divided by the saturation magnetization $M/M_{\mathrm{sat}}$ as a function of the field along the (110) crystal direction is also shown in panel E (pink line, right axis), demonstrating the initial trend, $M/M_{\mathrm{sat}} \approx (\mathbf{q}_L - 1)$ and the concurrence of the jumps in magnetization with the abrupt changes in the position of elastic scattering. (G) The intensity of the diffraction peaks in parts B-F. The AFM order that emerges for $\mu_0 H>1~\rm T$ is considerably weaker than the low field order, while the low field order falls off very abruptly for $\mu_0 H>1.2$~T.  Symbols are the same as in panels E and F.  (H) The zero field magnetic structure and dipole interactions in \YPP. Yb moments in five unit cells are shown along the diagonal, $(110)$ direction, with the interchain couplings, $\mathcal{J}^{\perp}_n$, as indicated.  The Yb SSL AFM layers are highlighted (blue and green arrows), with the periodicity given by FM pairs every five unit cells (black and red arrows).}
\end{figure}

The longitudinal interchain mode changes dramatically over a relatively narrow range of fields as the underlying antiferromagnetic order is weakened and ultimately destroyed. The low temperature $\mathbf{H}-T$ magnetic phase diagram of \YPP\ [Fig.~\ref{InField2}(A)] has several different AFM ordered phases \cite{Ochiai_JPSJ_2011, Shimura_JPSJ_2012, Iwakawa_JPSJ_2012, Kim_PRL_2013}. In zero field, there is a five by five periodicity to the order in the tetragonal $\mathbf{a}-\mathbf{b}$ plane, evidenced by neutron diffraction peaks that index as \qh$=0.2~\mathrm{rlu}$ [Fig.~\ref{Diffraction}(A)]~\cite{Miiller_PRB_2016}. When the gap $\Delta_\mathrm{S}$ closes at $\mu_0 H= 0.5~\rm T$, those peaks move from \ql$=1~\mathrm{rlu}$ to incommensurate positions along \ql\ [Fig.~\ref{Diffraction}(B)], consistent with the longitudinal component of the spin-spin correlation function probed by our neutron scattering measurements being locked to twice the Fermi wave vector~\cite{Giamarchi_book_2004}. The ordering wave vector follows 2\kf\ in turn, connecting to the softest parts of the excitation spectrum. There are several small and abrupt shifts in the ordering wave vector along both \ql\ and \qh\ [Fig.~\ref{Diffraction}(A-F)] that coincide with abrupt jumps in the derivative of the low temperature magnetization [Figs~\ref{InField1}(A) and~\ref{Diffraction}(E, F)], which manifest changes in 3D magnetic ordering as the magnetic moments re-arrange to minimize the energy of magnetic dipole interactions. For fields $\mu_0 H>1.0~\rm T$, there is an emergence of a second incommensurate AFM ordered phase that is  accompanied by the swift collapse of the original five by five order for $\mu_0 H>1.2~\rm T$ and even a third incommensurate order that persists up to the saturation field [Fig.~\ref{Diffraction}(C-G)]. As the new longitudinal interchain mode is an excitation of the underlying order, it is not surprising that it changes so dramatically between 1.0 and 1.5 T.

For $0.5 \lesssim \mu_0 H \lesssim 1$~T, the \ql\ component of the magnetic Bragg peak follows the magnetization [Fig.~\ref{Diffraction}E], which reflects the position of the spinon Fermi wavevector, 2\kf\ [Fig.~\ref{InField1}(B)] (See Supplementary Note 6). Theoretical description of the excitation spectrum in this spin-density-wave (SDW) phase can be obtained by applying bosonization methods for quasi-1D spin-1/2 antiferromagnets \cite{Schulz_PRL_1996,Essler_PRB_1997}. The low energy sector of such an antiferromagnet is described by the theory of noninteracting bosonic field $\phi$ governed by the Lagrangian, 
\begin{equation}
L = \frac{1}{4\pi}\int dx \Big[ v^{-1}(\partial_{\tau}\phi)^2 + v(\partial_x\phi)^2\Big], \label{L}
\end{equation}
where $v \sim J/a$. The expression for the $z$-component of the spins is
\begin{equation}
S^z - M = \frac{1}{2\pi}\partial_x\phi + A\sin[\phi + \pi(1- 2M)x], ~~ M = \langle S^z \rangle, \label{Sz}
\end{equation}
where $A$ is an amplitude with dimension of inverse length. A quantitative description of the interchain dispersion requires knowledge of the relevant couplings for the underlying order. In \YPP, the interchain interaction is predominantly of dipole-dipole origin. The symmetry of the Yb lattice sites suppresses magnetic dipole interaction between the two orthogonal sublattices, which cancels on the mean field level.  On the other hand, due to the Ising nature of Yb magnetic moments the intra-sublattice interactions $\mathcal{J}^{\perp}_{\mathrm{intra}}$ involve only $z$-components of effective spins-1/2 [Fig.~\ref{Diffraction}(H)]. Hence, the interchain coupling can be written as, ${\cal J}_{m}^{\perp}\int dx S^z(n,x)S^z(m+n,x)$, where $n,n+ m$ label different chains. Using (\ref{Sz}) and neglecting the marginal terms with derivatives of $\phi$ we get the following contribution to (\ref{L}),
 \begin{equation}
 L' = \sum_{n,m} A^2{\cal J}_{m}^{\perp}\int dx \cos[\phi(n,x) - \phi(m +n,x)].
 \end{equation}
Since fluctuations in 3D do not lead to divergencies, we can expand cosines in $L'$ around its minimum and get the Lagrangian quadratic in $\phi$. The zero field magnetic structure in \YPP\ suggests couplings up to fifth neighbors in the basal plane. We thus obtain a gapless longitudinal ``phason'' mode with the interchain dispersion,  
\begin{equation}
\label{E_{mode}}
E_{\mathrm{mode}}\left(\mathbf{q}_{HH}\right)^2 = 2\sum_{n=1}^{5} {\cal J}_n^{\perp}[1- \cos(2\pi n\mathbf{q}_{HH})],
\end{equation}
which accurately describes the 1.0~T data in Fig.~\ref{InField2}B.  

The situation is different at 1~T~$< \mu_0H <$~2.2~T (region marked by green circles on the phase diagram, Fig.~\ref{InField2}A). In this region, the static susceptibility is nonzero and magnetization smoothly increases, but the longitudinal interchain mode is gapped and magnetic Bragg peaks are locked to $q_L \approx 0.29$~rlu and do not change their $q_L$ positions with field [Fig.~\ref{Diffraction}E]. In principle, this presents a puzzle, which could be resolved by assuming that the transverse components of the effective spins order in a spiral configuration, while the $z$-component has a field independent ordering wave vector. Since neutrons do not register transverse fluctuations in \YPP, the corresponding Goldstone mode is invisible. In order to describe the gapped longitudinal mode observed in this phase, we add a phenomenological gap term, $\Delta_\mathrm{M}^2$, to Eq.~\ref{E_{mode}}, resulting in a dispersion, $E_{\mathrm{mode}}\left(\mathbf{q}_{HH}\right)^2 = \Delta_\mathrm{M}^2\left(1-\frac{1}{3\mathcal{J}^{\perp}_{\mathrm{tot}}} \sum_{n=1}^5\mathcal{J}^{\perp}_n \cos\left(2\pi n \mathbf{q}_{HH}\right)\right)$, where we normalize the couplings to the total interchain exchange, $\mathcal{J}^{\perp}_{\mathrm{tot}}=\sum_{n=1}^5\mathcal{J}^{\perp}_n$. We fit this expression to the measured modes by varying the relative couplings $\mathcal{J}^{\perp}_n/\mathcal{J}^{\perp}_{\mathrm{tot}}$, resulting in excellent agreement at all fields [Fig.~\ref{InField2}(B-D)]. Increasing the field from 1.0 to 1.5 T dramatically changes the relative interchain coupling strengths. The nearest neighbor coupling is reduced by approximately a factor of 11 when the field is increased from 1.0 to 1.5 T, while the next nearest neighbor coupling is reduced by a factor of 2 and changes sign from ferromagnetic to antiferromagnetic. Smaller changes are found in the higher order terms. These changes are consistent with the observation from the magnetic diffraction [Fig.~\ref{Diffraction}] of a tendency towards a weaker, frustrated longitudinal antiferromagnetic order as the applied magnetic field progressively polarizes the moments (See Supplementary Note 7).

It is difficult to visualize the nature of the interchain mode itself, as it is purely quantum mechanical in its origin, with no simple classical analog. In a conventional 3D ordered magnet, these interchain excitations would be transverse spin waves -- pseudo-Goldstone modes of an antiferrromagneic order parameter, acquiring a small gap (mass) in the presence of spin anisotropy. The longitudinal polarization reveals that the new excitations observed here in \YPP, which are separated from $E=0$ with a field-dependent gap $\delta < 0.12~\mathrm{meV}$, are in fact far more exotic. They represent amplitude excitations of the AFM order parameter, {\it ie} the staggered magnetization~\cite{Ruegg_PRL_2008, Merchant_NatPhys_2014}, analogous to the amplitude modes of the superconducting order parameter found in NbSe$_2$~\cite{Sooryakumar_PRL_1980, Sooryakumar_PRB_1981} and the Higgs boson \cite{Pekker_ARCMP_2015} itself. There is a large literature on the subject in the context of the theory of quantum magnets ({\it eg} \cite{Schulz_PRL_1996, Essler_PRB_1997, Affleck_PRL_1989, Normand_PRB_1997}), but so far there have been few experiments among weakly coupled chain systems that probe this mode and its dispersion across different regimes of interchain coupling, or its decay into transverse magnons in detail \cite{Affleck_PRB_1992}. This damping causes the longitudinal mode to appear as a resonance in the longitudinally polarized continuum rather than the dispersing quasiparticle like excitation observed here in \YPP\ \cite{Lake_NatMat_2005,Lake_PRL_2000}, which often obscures the physics entirely and has even led some to question the assumptions behind the theory~\cite{Zheludev_PRL_2002}. Recently, some evidence for a sharp longitudinal mode coexisting with the transverse spin waves has been obtained via polarized neutron scattering measurements in a more strongly coupled 2D ladder system \cite{Hong_NatPhys_2017}, while other materials with similar \textit{XXZ} anisotropy to \YPP\ tend to confine all spinons into many modes~\cite{Grenier_PRL_2015, Matsuda_PRB_2017, Faure_NatPhys_2018, Wang_PRB_2015, Bera_PRB_2017}.  Here we overcome the limitations of such experiments thanks to the tuning parameters of \YPP, which uniquely single out the longitudinal channel and allow us to clearly identify the dispersing amplitude mode and the deconfined 1D excitations at higher energies.

The physics of \YPP\ that suppresses transverse spin waves also protects these longitudinal excitations and allows detailed observation of this mode dispersion and its dependence on an applied magnetic field, which tunes the ordered state across different phases. Accompanying transverse excitations -- the spin waves -- must exist, but are not observed [Fig.~\ref{InField2}(F)]. They are suppressed by a factor ($g_{||}/g_{\perp})^2 \gtrsim 100$ and can not be measured in this scattering geometry, where even the application of a substantial magnetic field leaves behind the longitudinal continuum from magnetic moments perpendicular to the field. While further experiments are needed in alternate scattering geometries to clearly observe the accompanying transverse excitations, the present results quantify in unique detail an unusual dispersing longitudinal mode, a Higgs-like excitation of a effective spin-1/2 order parameter across different ordered phases induced by magnetic fields in \YPP.

\section*{Methods}

\subsection*{Neutron Scattering}
The \YPP\ sample used in these measurements was the same sample used in Ref.~\cite{Wu_Science_2016}.  The sample consists of approximately 400 co-aligned \YPP\ single crystals (total mass $\approx 6~\rm g$) mounted to aluminum plates.  For all measurements  the sample was oriented with the (1{\=1}0) crystal direction vertical leaving the (110) and (001) directions in the horizontal scattering plane, making the scattering plane $(H, H, L)$ in reciprocal space.  All momenta are given in reciprocal lattice units (rlu), with 1 rlu given by $2\pi/a=2\pi/7.76~\rm{\AA}=0.810~\rm{\AA}^{-1}$ along the $H$ direction and $2\pi/c=2\pi/7.02~\rm{\AA}=0.895~\rm{\AA}^{-1}$ along $L$.  The crystallographic unit cell is twice the Yb-Yb near neighbor spacing along the $c$-axis -- the relevant spacing for spinons.  Therefore, the Brillouin zone for spinons is indexed from 0 to 2 rlu along $(0, 0, L)$, rather than the typical 0 to 1 rlu.  Notationally, \qh\ is parallel to the $(H, H, 0)$ direction, with scattering primarily coming from  $(H, H, 1)$ while \ql\ is along the $(0, 0, L)$ direction.

The inelastic neutron scattering measurements on \YPP\ making up the bulk of the data in this paper were made on the OSIRIS spectrometer at the ISIS neutron source at Rutherford Appleton Laboratory in Didcot, Oxfordshire, UK~\cite{Demmel_NIMPRA_2014}.  The sample was mounted in a dilution refrigerator inside of a 7 T superconducting magnet with the field in the vertical direction parallel to the (1{\=1}0) crystal direction.  OSIRIS is an indirect geometry, time-of-flight neutron spectrometer.  The incident neutron beam is a white beam of cold neutrons.  The PG002 analyzer was used with a final neutron energy $E_f=1.84~{\rm meV}$ ($\lambda=6.67~{\rm \AA}$).  The energy resolution was $\approx 0.03~\mathrm{meV}$ at $E=0~\mathrm{meV}$.  Due to the magnet construction, only scattering into a range of angles $45^{\circ}<2\theta<135^{\circ}$ is permitted.  We therefore discard the 14 detector channels at the lowest scattering angles and the 3 channels at the highest scattering angles, determined by measurements of a vanadium standard in the magnet.  All data were corrected for the Yb$^{3+}$ form factor~\cite{Brown_Xtal_book}.

The measurements on OSIRIS were made at $T=0.140~\rm K$.  A field greater than 2.3 T along the (1{\=1}0) crystal direction polarizes all of the magnetic moments parallel to that direction, while leaving the orthogonal moments along the (110) direction in the horizontal plane unaffected due to the ground state doublet of the Yb ions~\cite{Wu_Science_2016}.  Measurements made at $\mu_0 H=4~\rm T$ can therefore be used as a background for measurements made at $\mu_0 H<2.3~\rm T$, isolating only the lower field scattering contribution from magnetic moments oriented parallel to the field.  All neutron scattering results from OSIRIS have a measurement made at $T=0.140~\rm K$ and $\mu_0 H=4~\rm T$ subtracted in this fashion.  For the nominal zero field measurements, a small bias field of 0.025 T was used to suppress superconductivity in the aluminum sample holder.

In general, inelastic neutron scattering probes the dynamical spin correlation function~\cite{Squires_book_1978}.  Because of the crystal field ground state doublet of the Yb ions in \YPP\ we are sensitive only to the longitudinal component of this function (see Supplementary Note 5).

Because our detector coverage does not include an entire Brillouin zone, ($0<\mathbf{q}_L<2~{\rm rlu}$, $0<\mathbf{q}_{HH}<1~{\rm rlu}$) we integrate all data in the scattering plane in the reciprocal space direction orthogonal to the direction being considered.  The \ql\ dependencies show in Figs.~\ref{ZeroField}(B) and~\ref{InField1}(D-E) (And Supplementary Fig. 1) integrate the entire measured range of \qh.  The same \qh\ integration was used to demonstrate the amount of low energy spectral weight in the bound state (main text Fig.~\ref{InField2}(E) inset).  Similarly, all \ql\ were integrated to examine the \qh\ dependence in Figs.~\ref{ZeroField}(C) and~\ref{InField2}(B-D) (and Supplementary Figs. 4 and 5) and in the cuts used to extract the interchain mode dispersion in Fig.~\ref{InField2}(E).  Although our data from OSIRIS do not cover an entire Brillouin zone, the symmetry of our measurements about both $\mathbf{q}_L=1~{\rm rlu}$ and $\mathbf{q}_{HH}=0~{\rm rlu}$ confirms that our coverage is sufficient.

Inelastic scattering measurements of the polarization factor (main text, Fig. 3(F)) and diffraction measurements (main text, Fig.~4(B-H))  were performed on the Disk Chopper Spectrometer (DCS) at the Center for Neutron Research at the National Institute for Standards and Technology in Gaithersburg, MD, USA~\cite{Copley_CP_2003}.  For these measurements, the same sample was mounted in the same scattering geometry as the OSIRIS experiments, inside of a dilution refrigerator inside of a 10 T vertical superconducting magnet.  DCS is a direct geometry spectrometer and $E_i=3.27~{\rm meV}$ ($\lambda=5.00~{\rm \AA}$) was used.  The energy resolution was $\approx 0.1~\mathrm{meV}$.  The temperature of these measurements was $T=0.07~\rm K$.  Measurements on DCS have a similarly measured background subtracted and the scattering was corrected for the Yb$^{3+}$ form factor~\cite{Brown_Xtal_book} and neutron absorption using the DAVE software package~\cite{Azuah_JRN_2009} in the same manner described in the Supplementary Materials for~\cite{Wu_Science_2016}.  The measurement at $\mu_0 H=4~\rm T$ of the polarization factor (main text Fig. 3(F)) is simply the $\mu_0 H=4~{\rm T}$ measurement on its own, with no background subtracted.  For the nominal zero field measurements, a small bias field of 0.025 T was used to suppress superconductivity in the aluminum sample holder.

\subsection*{Specific Heat and Magnetization}
Specific heat measurements used in the phase diagram shown in Fig.~\ref{Diffraction}(A) (and Supplementary Figs. 2 and 3) were made using a Physical Property Measurement System (PPMS) made by Quantum Designs on a single crystal of \YPP\ (note: the identification of the equipment used in the various measurements is not intended to imply recommendation or endorsement by the National Institute of Standards and Technology, nor is it intended to imply that this equipment is necessarily the best available for the purpose.)  Measurements for $T < 1.8~\mathrm{K}$ utilized a PPMS dilution refrigerator insert.  The (110) direction was oriented vertically, parallel to the magnetic field.  Magnetization measurements for $T<1.8$ K in Figs.~\ref{InField1}(A) and~\ref{Diffraction}(A,F) (and Supplementary Fig. 6) were made using a PPMS Hall sensor magnetometer in a dilution refrigerator insert at $T=0.150~\rm K$.  Magnetization measurements for $T>1.8$ K in Fig.~\ref{Diffraction}(A) were made using a Quantum Designs Magnetic Property Measurement System SQUID magnetometer.  The magnetic field for both sets of measurements was along the (110) direction.

\subsection*{Obtaining the Position of the Interchain mode}
The interchain mode dispersions shown in Figs.~\ref{InField2}(B-D) (and Supplementary Fig. 5) were extracted by integrating a window of $\mathbf{q}_{HH} = 0.1~{\rm rlu}$ and fitting the resulting cuts of the scattering function as a function of energy $S\left(E\right)$, examples of which are shown in Fig.~\ref{InField2}(E) of the main text.  The fitting function  represents the mode as a damped harmonic oscillator -- the product of the Bose population factor with the difference of two Lorentzians, centered at positive and negative energies $E_{\mathrm{mode}}$, with a width $\Gamma$ and amplitude $A$, Eqn.~\ref{DHO}.
\begin{equation}\label{DHO}
I\left(E\right) = \frac{A}{1-e^{-E/k_\mathrm{B}T}}\left(\frac{\Gamma}{\left(E-E_{\mathrm{mode}}\right)^2+\Gamma^2}-\frac{\Gamma}{\left(E+E_{\mathrm{mode}}\right)^2+\Gamma^2}\right)
\end{equation}
This function was added to the product of another Lorentzian and step function each centered at energy $E>E_{\mathrm{mode}}$ representing the continuum. The fit itself was  performed to these two functions convolved with a gaussian representing the instrumental resolution.

\subsection*{Time-Dependent Density Matrix Renormalization Group Calculations}
The longitudinal component of the dynamical spin structure factor $S(\mathbf{q},\omega)$ was obtained by means of the time-dependent density matrix renormalization group (tDMRG) method
\cite{White2004,Daley2004,Feiguin2011vietri,Feiguin2013a}. The approach has been extensively described in the literature and essentially consists of calculating the time-dependent correlation function:
\begin{equation}
S(x-x_0,t)=\langle \psi|e^{i\hat{H}t}\hat S^z(x)e^{-i\hat{H}t}\hat S^z(x_0)|\psi\rangle.
\label{gf}
\end{equation}
The operator $\hat S^z(x_0=L/2)$ is applied at the center of the chain and the resulting state is evolved in real-time. At every step, the overlap with the state $\hat S^z(x)e^{-i\hat{H}t}|\psi\rangle$ is measured and the correlations function in real time and space is recorded. The results are Fourier transformed to frequency and momentum using a properly chosen Hann window that determines the resolution of the final result. In our case, in order to compare to experiments we have used a window of half-width $\Delta t=7$ in units of $1/J$. Only two things differ from conventional calculations: since the parameters considered fall into the Ising phase of the model, which tends to break translational symmetry at finite magnetization, the simulations were carried out using periodic boundary conditions and the time-targeting method with a Krylov expansion of the evolution operator~\cite{Feiguin2005}. In addition, we plot the results for the operator $(\hat S^z(x) - m/2)$ (instead of $\hat S^z(x)$), where $m$ is the magnetization value. The offset
allows us to resolve the density fluctuations and eliminates large contributions at low frequencies and momenta~\cite{Nocera2016}. Surprisingly, due to the low entanglement in the Ising phase, truncation errors smaller than $10^{-5}$ are easily achievable using 300 states, even on a chain of length $L=96$ with periodic boundary conditions.

The color scale for the calculations displayed in Figs. 1(D) and 2(G-I) was determined by normalizing the integral of the $E>0$ portion of the calculation to the measured integral of the inelastic intensity at the the same field.

\bibliographystyle{naturemag}

\begin{thebibliography}{10}

\bibitem{Bethe_Zphys_1931}
\bibinfo{author}{Bethe, H.}
\newblock \bibinfo{title}{Zur theorie der metalle i. eigenwerte und
  eigenfunktionen der linearen atomkette}.
\newblock \emph{\bibinfo{journal}{Zeitschrift f\"ur Physik}}
  \textbf{\bibinfo{volume}{71}}, \bibinfo{pages}{205--226}
  (\bibinfo{year}{1931}).

\bibitem{Haldane_PRL1980}
\bibinfo{author}{Haldane, F. D.~M.}
\newblock \bibinfo{title}{General relation of correlation exponents and
  spectral properties of one-dimensional fermi systems: Application to the
  anisotropic $s=\frac{1}{2}$ heisenberg chain}.
\newblock \emph{\bibinfo{journal}{Physical Review Letters}}
  \textbf{\bibinfo{volume}{45}}, \bibinfo{pages}{1358--1362}
  (\bibinfo{year}{1980}).


\bibitem{Fadeev_PhysLett1981}
\bibinfo{author}{Fadeev, L.~D.} \& \bibinfo{author}{Takhtajan, L.~A.}
\newblock \bibinfo{title}{What is the spin of a spin wave?}
\newblock \emph{\bibinfo{journal}{Physics Letters A}}
  \textbf{\bibinfo{volume}{85}}, \bibinfo{pages}{375--377}
  (\bibinfo{year}{1981}).

\bibitem{Polyakov_NucPhys1977}
\bibinfo{author}{Polyakov, A.}
\newblock \bibinfo{title}{Quark confinement and topology of gauge theories}.
\newblock \emph{\bibinfo{journal}{Nuclear Physics B}}
  \textbf{\bibinfo{volume}{120}}, \bibinfo{pages}{429--458}
  (\bibinfo{year}{1977}).

\bibitem{Shelton_PRB_1996}
\bibinfo{author}{Shelton, D.~G.}, \bibinfo{author}{Nersesyan, A.~A.} \&
  \bibinfo{author}{Tsvelik, A.~M.}
\newblock \bibinfo{title}{Antiferromagnetic spin ladders: Crossover between
  spin $s=1/2$ and $s=1$ chains}.
\newblock \emph{\bibinfo{journal}{Physical Review B}}
  \textbf{\bibinfo{volume}{53}}, \bibinfo{pages}{8521--8532}
  (\bibinfo{year}{1996}).

\bibitem{Lake_NatMat_2005}
\bibinfo{author}{Lake, B.}, \bibinfo{author}{Tennant, D.~A.},
  \bibinfo{author}{Frost, C.~D.} \& \bibinfo{author}{Nagler, S.~E.}
\newblock \bibinfo{title}{Quantum criticality and universal scaling of a
  quantum antiferromagnet}.
\newblock \emph{\bibinfo{journal}{Nature Materials}}
  \textbf{\bibinfo{volume}{4}}, \bibinfo{pages}{329--334}
  (\bibinfo{year}{2005}).

\bibitem{Zaliznyak_NatMat2005}
\bibinfo{author}{Zaliznyak, I.~A.}
\newblock \bibinfo{title}{A glimpse of a luttinger liquid}.
\newblock \emph{\bibinfo{journal}{Nature Materials}}
  \textbf{\bibinfo{volume}{4}}, \bibinfo{pages}{273--275}
  (\bibinfo{year}{2005}).

\bibitem{Lake_NatPhys_2009}
\bibinfo{author}{Lake, B.} \emph{et~al.}
\newblock \bibinfo{title}{Confinement of fractional quantum number particles in
  a condensed-matter system}.
\newblock \emph{\bibinfo{journal}{Nature Physics}}
  \textbf{\bibinfo{volume}{6}}, \bibinfo{pages}{50--55} (\bibinfo{year}{2009}).

\bibitem{Tsvelik_book_2003}
\bibinfo{author}{Tsvelik, A.~M.}
\newblock \emph{\bibinfo{title}{Quantum Field Theory in Condensed Matter
  Physics}} (\bibinfo{publisher}{Cambridge University Press},
  \bibinfo{year}{2003}), \bibinfo{edition}{second} edn.

\bibitem{Giamarchi_book_2004}
\bibinfo{author}{Giamarchi, T.}
\newblock \emph{\bibinfo{title}{Quantum Physics in One Dimension}}
  (\bibinfo{publisher}{Oxford Science Publications}, \bibinfo{year}{2004}).

\bibitem{Grenier_PRL_2015}
\bibinfo{author}{Grenier, B.} \emph{et~al.}
\newblock \bibinfo{title}{Longitudinal and transverse zeeman ladders in the
  ising-like chain antiferromagnet
  ${\mathrm{baco}}_{2}{\mathrm{v}}_{2}{\mathrm{o}}_{8}$}.
\newblock \emph{\bibinfo{journal}{Physical Review Letters}}
  \textbf{\bibinfo{volume}{114}}, \bibinfo{pages}{017201(1--5)}
  (\bibinfo{year}{2015}).

\bibitem{Matsuda_PRB_2017}
\bibinfo{author}{Matsuda, M.} \emph{et~al.}
\newblock \bibinfo{title}{Magnetic structure and dispersion relation of the
  $s=\frac{1}{2}$ quasi-one-dimensional ising-like antiferromagnet
  ${\mathrm{baco}}_{2}{\mathrm{v}}_{2}{\mathrm{o}}_{8}$ in a transverse
  magnetic field}.
\newblock \emph{\bibinfo{journal}{Physical Review B}}
  \textbf{\bibinfo{volume}{96}}, \bibinfo{pages}{024439(1--8)}
  (\bibinfo{year}{2017}).

\bibitem{Faure_NatPhys_2018}
\bibinfo{author}{Faure, Q.} \emph{et~al.}
\newblock \bibinfo{title}{Topological quantum phase transition in the
  $\mathrm{I}$sing-like antiferromagnetic spin chain $\mathrm{BaCo_2V_2O_8}$}.
\newblock \emph{\bibinfo{journal}{Nature Physics}}
  \textbf{\bibinfo{volume}{14}}, \bibinfo{pages}{716--723}
  (\bibinfo{year}{2018}).

\bibitem{Wang_PRB_2015}
\bibinfo{author}{Wang, Z.} \emph{et~al.}
\newblock \bibinfo{title}{Spinon confinement in the one-dimensional
  $\mathrm{I}$sing-like antiferromagnet $\mathrm{SrCo_2V_2O_8}$}.
\newblock \emph{\bibinfo{journal}{Physical Review B}}
  \textbf{\bibinfo{volume}{91}}, \bibinfo{pages}{140404(R)(1--4)}
  (\bibinfo{year}{2015}).

\bibitem{Bera_PRB_2017}
\bibinfo{author}{Bera, A.~K.} \emph{et~al.}
\newblock \bibinfo{title}{Spinon confinement in a quasi-one-dimensional
  anisotropic heisenberg magnet}.
\newblock \emph{\bibinfo{journal}{Physical Review B}}
  \textbf{\bibinfo{volume}{96}}, \bibinfo{pages}{054423(1--17)}
  (\bibinfo{year}{2017}).

\bibitem{Ruegg_PRL_2008}
\bibinfo{author}{Ru\"egg, C.} \emph{et~al.}
\newblock \bibinfo{title}{Quantum magnets under pressure: Controlling
  elementary excitations in $\mathrm{TlCuCl}_3$}.
\newblock \emph{\bibinfo{journal}{Physical Review Letters}}
  \textbf{\bibinfo{volume}{100}}, \bibinfo{pages}{205701(1--4)}
  (\bibinfo{year}{2008}).

\bibitem{Merchant_NatPhys_2014}
\bibinfo{author}{Merchant, P.} \emph{et~al.}
\newblock \bibinfo{title}{Quantum and classical criticality in a dimerized
  quantum antiferromagnet}.
\newblock \emph{\bibinfo{journal}{Nature Physics}}
  \textbf{\bibinfo{volume}{10}}, \bibinfo{pages}{373--379}
  (\bibinfo{year}{2014}).

\bibitem{Pekker_ARCMP_2015}
\bibinfo{author}{Pekker, D.} \& \bibinfo{author}{Varma, C.~M.}
\newblock \bibinfo{title}{Amplitude/${\rm higgs}$ modes in condensed matter
  physics}.
\newblock \emph{\bibinfo{journal}{Annual Review of Condensed Matter Physics}}
  \textbf{\bibinfo{volume}{6}}, \bibinfo{pages}{269--297}
  (\bibinfo{year}{2015}).

\bibitem{Shastry_PB_1981}
\bibinfo{author}{Shastry, B.~S.} \& \bibinfo{author}{Sutherland, B.}
\newblock \bibinfo{title}{Exact ground state of a quantum mechanical
  antiferromagnet}.
\newblock \emph{\bibinfo{journal}{Physica B}} \textbf{\bibinfo{volume}{108}},
  \bibinfo{pages}{1069--1070} (\bibinfo{year}{1981}).

\bibitem{Pottgen_JSSC_1999}
\bibinfo{author}{P\"ottgen, R.} \emph{et~al.}
\newblock \bibinfo{title}{Structure and properties of $\mathrm{YbZnSn}$,
  $\mathrm{YbAgSn}$, and $\mathrm{Yb}_2\mathrm{Pt}_2\mathrm{Pb}$}.
\newblock \emph{\bibinfo{journal}{Journal of Solid State Chemistry}}
  \textbf{\bibinfo{volume}{145}}, \bibinfo{pages}{668--677}
  (\bibinfo{year}{1999}).

\bibitem{Kim_PRB_2008}
\bibinfo{author}{Kim, M.~S.}, \bibinfo{author}{Bennett, M.~C.} \&
  \bibinfo{author}{Aronson, M.~C.}
\newblock \bibinfo{title}{$\mathrm{Yb_2Pt_2Pb}$: Magnetic frustration in the
  shastry-sutherland lattice}.
\newblock \emph{\bibinfo{journal}{Physical Review B}}
  \textbf{\bibinfo{volume}{77}}, \bibinfo{pages}{144425(1--7)}
  (\bibinfo{year}{2008}).

\bibitem{Ochiai_JPSJ_2011}
\bibinfo{author}{Ochiai, A.} \emph{et~al.}
\newblock \bibinfo{title}{Field-induced partially disordered state in
  $\mathrm{Yb}_2\mathrm{Pt}_2\mathrm{Pb}$}.
\newblock \emph{\bibinfo{journal}{Journal of the Physical Society of Japan}}
  \textbf{\bibinfo{volume}{80}}, \bibinfo{pages}{123705(1--4)}
  (\bibinfo{year}{2011}).

\bibitem{Shimura_JPSJ_2012}
\bibinfo{author}{Shimura, Y.}, \bibinfo{author}{Sakakibara, T.},
  \bibinfo{author}{Iwakawa, K.}, \bibinfo{author}{Sugiyama, K.} \&
  \bibinfo{author}{Onuki, Y.}
\newblock \bibinfo{title}{Low temperature magnetization of
  $\mathrm{Yb_2Pt_2Pb}$ with the shastry-sutherland type lattice and a
  high-rank multipole interaction}.
\newblock \emph{\bibinfo{journal}{Journal of the Physical Society of Japan}}
  \textbf{\bibinfo{volume}{81}}, \bibinfo{pages}{103601(1--4)}
  (\bibinfo{year}{2012}).

\bibitem{Iwakawa_JPSJ_2012}
\bibinfo{author}{Iwakawa, K.} \emph{et~al.}
\newblock \bibinfo{title}{Multiple metamagnetic transitions in antiferromagnet
  $\mathrm{Yb_2Pt_2Pb}$ with the shastry-sutherland lattice}.
\newblock \emph{\bibinfo{journal}{Journal of the Physical Society of Japan}}
  \textbf{\bibinfo{volume}{81}}, \bibinfo{pages}{SB058(1--4)}
  (\bibinfo{year}{2012}).

\bibitem{Kim_PRL_2013}
\bibinfo{author}{Kim, M.~S.} \& \bibinfo{author}{Aronson, M.~C.}
\newblock \bibinfo{title}{Spin liquids and antiferromagnetic order in the
  shastry-sutherland-lattice compound $\mathrm{Yb_2Pt_2Pb}$}.
\newblock \emph{\bibinfo{journal}{Physical Review Letters}}
  \textbf{\bibinfo{volume}{110}}, \bibinfo{pages}{017201(1--6)}
  (\bibinfo{year}{2013}).

\bibitem{Wu_Science_2016}
\bibinfo{author}{Wu, L.~S.} \emph{et~al.}
\newblock \bibinfo{title}{Orital-exchange and fractional quantum number
  excitations in an f-electron metal, $\mathrm{Yb}_2\mathrm{Pt}_2\mathrm{Pb}$}.
\newblock \emph{\bibinfo{journal}{Science}} \textbf{\bibinfo{volume}{352}},
  \bibinfo{pages}{1206--1210} (\bibinfo{year}{2016}).

\bibitem{Squires_book_1978}
\bibinfo{author}{Squires, G.~L.}
\newblock \emph{\bibinfo{title}{Introduction to the Theory of Thermal Neutron
  Scattering}} (\bibinfo{publisher}{Cambridge University Press},
  \bibinfo{address}{England}, \bibinfo{year}{2012}), \bibinfo{edition}{third
  edition} edn.

\bibitem{White2004}
\bibinfo{author}{White, S.~R.} \& \bibinfo{author}{Feiguin, A.~E.}
\newblock \bibinfo{title}{Real-time evolution using the density matrix
  renormalization group}.
\newblock \emph{\bibinfo{journal}{Physical Review Letters}}
  \textbf{\bibinfo{volume}{93}}, \bibinfo{pages}{076401(1--4)}
  (\bibinfo{year}{2004}).

\bibitem{Miiller_PRB_2016}
\bibinfo{author}{Miiller, W.} \emph{et~al.}
\newblock \bibinfo{title}{Magnetic structure of
  $\mathrm{Yb}_2\mathrm{Pt}_2\mathrm{Pb}$: Ising moments on the
  shastry-sutherland lattice}.
\newblock \emph{\bibinfo{journal}{Physical Review B}}
  \textbf{\bibinfo{volume}{93}}, \bibinfo{pages}{104419(1--10)}
  (\bibinfo{year}{2016}).

\bibitem{Bougourzi_PRB1998}
\bibinfo{author}{Bougourzi, A.~H.}, \bibinfo{author}{Karbach, M.} \&
  \bibinfo{author}{M\"uller, G.}
\newblock \bibinfo{title}{Exact two-spinon dynamic structure factor of the
  one-dimensional $s=\frac{1}{2}$ heisenberg-ising antiferromagnet}.
\newblock \emph{\bibinfo{journal}{Physical Review B}}
  \textbf{\bibinfo{volume}{57}}, \bibinfo{pages}{11429--11438}
  (\bibinfo{year}{1998}).

\bibitem{Caux_JStatMech_2006}
\bibinfo{author}{Caux, J.~S.} \& \bibinfo{author}{Hagemans, R.}
\newblock \bibinfo{title}{The four-spinon dynamical structure factor of the
  heisenberg chain}.
\newblock \emph{\bibinfo{journal}{Journal of Statistical Mechanics: Theory and
  Experiment}} \textbf{\bibinfo{volume}{12}}, \bibinfo{pages}{P12013(1--14)}
  (\bibinfo{year}{2006}).

\bibitem{Greiner_book_1994}
\bibinfo{author}{Greiner, W.} \& \bibinfo{author}{Sch\"afer, A.}
\newblock \emph{\bibinfo{title}{Quantum Chromodynamics}}
  (\bibinfo{publisher}{Springer-Verlag Berlin-Hdidelberg},
  \bibinfo{year}{1994}).

\bibitem{Schulz_PRL_1996}
\bibinfo{author}{Schulz, H.~J.}
\newblock \bibinfo{title}{Dynamics of coupled quantum spin chains}.
\newblock \emph{\bibinfo{journal}{Physical Review Letters}}
  \textbf{\bibinfo{volume}{77}}, \bibinfo{pages}{2790--2793}
  (\bibinfo{year}{1996}).

\bibitem{Essler_PRB_1997}
\bibinfo{author}{Essler, F. H.~L.}, \bibinfo{author}{Tsvelik, A.~M.} \&
  \bibinfo{author}{Delfino, G.}
\newblock \bibinfo{title}{Quasi-one-dimensional spin-$\frac{1}{2}$ heisenberg
  magnets in their ordered phase: Correlation functions}.
\newblock \emph{\bibinfo{journal}{Physical Review B}}
  \textbf{\bibinfo{volume}{56}}, \bibinfo{pages}{11001--11013}
  (\bibinfo{year}{1997}).

\bibitem{Sooryakumar_PRL_1980}
\bibinfo{author}{Sooryakumar, R.} \& \bibinfo{author}{Klein, M.~V.}
\newblock \bibinfo{title}{Raman scattering by superconducting-gap excitations
  and their coupling to charge-density waves}.
\newblock \emph{\bibinfo{journal}{Physical Review Letters}}
  \textbf{\bibinfo{volume}{45}} (\bibinfo{year}{1980}).

\bibitem{Sooryakumar_PRB_1981}
\bibinfo{author}{Sooryakumar, R.} \& \bibinfo{author}{Klein, M.~V.}
\newblock \bibinfo{title}{Raman scattering from superconducting gap excitations
  in the presence of a magnetic field}.
\newblock \emph{\bibinfo{journal}{Physical Review B}}
  \textbf{\bibinfo{volume}{23}}, \bibinfo{pages}{3213--3221}
  (\bibinfo{year}{1981}).

\bibitem{Affleck_PRL_1989}
\bibinfo{author}{Affleck, I.}
\newblock \bibinfo{title}{Model for quasi-one-dimensional antiferromagnets:
  Application to $\mathrm{CsNiCl}_3$}.
\newblock \emph{\bibinfo{journal}{Physical Review Letters}}
  \textbf{\bibinfo{volume}{62}}, \bibinfo{pages}{474--477}
  (\bibinfo{year}{1989}).

\bibitem{Normand_PRB_1997}
\bibinfo{author}{Normand, B.} \& \bibinfo{author}{Rice, T.~M.}
\newblock \bibinfo{title}{Dynamical properties of an antiferromagnet near the
  quantum critical point: Application to $\mathrm{LaCuO}_{2.5}$}.
\newblock \emph{\bibinfo{journal}{Physical Review B}}
  \textbf{\bibinfo{volume}{56}}, \bibinfo{pages}{8760--8773}
  (\bibinfo{year}{1997}).

\bibitem{Affleck_PRB_1992}
\bibinfo{author}{Affleck, I.} \& \bibinfo{author}{Wellman, G.~F.}
\newblock \bibinfo{title}{Longitudinal modes in quasi-one-dimensional
  antiferromagnets}.
\newblock \emph{\bibinfo{journal}{Physical Review B}}
  \textbf{\bibinfo{volume}{46}}, \bibinfo{pages}{8934--8953}
  (\bibinfo{year}{1992}).

\bibitem{Lake_PRL_2000}
\bibinfo{author}{Lake, B.}, \bibinfo{author}{Tennant, D.~A.} \&
  \bibinfo{author}{Nagler, S.~E.}
\newblock \bibinfo{title}{Novel longitudinal mode in the coupled quantum chain
  compound $\mathrm{KCuF}_3$}.
\newblock \emph{\bibinfo{journal}{Physical Review Letters}}
  \textbf{\bibinfo{volume}{85}}, \bibinfo{pages}{832--835}
  (\bibinfo{year}{2000}).

\bibitem{Zheludev_PRL_2002}
\bibinfo{author}{Zheludev, A.}, \bibinfo{author}{Kakurai, K.},
  \bibinfo{author}{Matsuda, T.}, \bibinfo{author}{Uchinokura, K.} \&
  \bibinfo{author}{Nakajima, K.}
\newblock \bibinfo{title}{Dominance of the excitation continuum in the
  longitudinal spectrum of weakly coupled heisenberg $s=1/2$ chains}.
\newblock \emph{\bibinfo{journal}{Physical Review Letters}}
  \textbf{\bibinfo{volume}{89}}, \bibinfo{pages}{197205(1--4)}
  (\bibinfo{year}{2002}).

\bibitem{Hong_NatPhys_2017}
\bibinfo{author}{Hong, T.} \emph{et~al.}
\newblock \bibinfo{title}{Higgs amplitude mode in a two-dimensional quantum
  antiferromagnet near the quantum critical point}.
\newblock \emph{\bibinfo{journal}{Nature Physics}}
  \textbf{\bibinfo{volume}{13}}, \bibinfo{pages}{638--643}
  (\bibinfo{year}{2017}).

\bibitem{Demmel_NIMPRA_2014}
\bibinfo{author}{Demmel, F.} \& \bibinfo{author}{Pokhilchuk, K.}
\newblock \bibinfo{title}{The resolution of the tof-backscattering spectrometer
  osiris: Monte carlo simulations and analytical calculations}.
\newblock \emph{\bibinfo{journal}{Nuclear Instruments and Methods in Physics
  Research A}} \textbf{\bibinfo{volume}{767}}, \bibinfo{pages}{426--432}
  (\bibinfo{year}{2014}).

\bibitem{Brown_Xtal_book}
\bibinfo{author}{Brown, P.~J.}
\newblock \emph{\bibinfo{title}{International Tables for Crystallography}},
  vol. \bibinfo{volume}{C, Parts 4 and 6} (\bibinfo{publisher}{Springer},
  \bibinfo{year}{2006}).

\bibitem{Copley_CP_2003}
\bibinfo{author}{Copley, J. R.~D.} \& \bibinfo{author}{Cook, J.~C.}
\newblock \bibinfo{title}{The disk chopper spectrometer at nist: a new
  instrument for quasielastic neutron scattering studies}.
\newblock \emph{\bibinfo{journal}{Chemical Phyics}}
  \textbf{\bibinfo{volume}{292}}, \bibinfo{pages}{477--485}
  (\bibinfo{year}{2003}).

\bibitem{Azuah_JRN_2009}
\bibinfo{author}{Azuah, R.~T.} \emph{et~al.}
\newblock \bibinfo{title}{Dave: A comprehensive software suite for the
  reduction, visualization, and analysis of low energy neutron spectroscopic
  data}.
\newblock \emph{\bibinfo{journal}{Journal of Research of the National Institute
  of Standards and Technology}} \textbf{\bibinfo{volume}{114}},
  \bibinfo{pages}{341--358} (\bibinfo{year}{2009}).

\bibitem{Daley2004}
\bibinfo{author}{Daley, A.}, \bibinfo{author}{Kollath, C.},
  \bibinfo{author}{Schollw\"ock, U.} \& \bibinfo{author}{Vidal, G.}
\newblock \bibinfo{title}{Time-dependent density matrix renormalization using
  adapive effective hilbert spaces}.
\newblock \emph{\bibinfo{journal}{Journal of Statistical Mechanics: Theory and
  Experiment}} \bibinfo{pages}{P04005(1--28)} (\bibinfo{year}{2004}).

\bibitem{Feiguin2011vietri}
\bibinfo{author}{Feiguin, A.~E.}
\newblock \bibinfo{title}{The density matrix renormalization group and its time
  dependent variants}.
\newblock In \bibinfo{editor}{Avella, A.} \& \bibinfo{editor}{Mancini, F.}
  (eds.) \emph{\bibinfo{booktitle}{Lectures on the Physics of strongly
  correlated systems XV}} (\bibinfo{publisher}{AIP}, \bibinfo{year}{2011}).

\bibitem{Feiguin2013a}
\bibinfo{author}{Feiguin, A.~E.}
\newblock \bibinfo{title}{The time-dependent density matrix renormalization
  group}.
\newblock In \bibinfo{editor}{Avella, A.} \& \bibinfo{editor}{Mancini, F.}
  (eds.) \emph{\bibinfo{booktitle}{Strongly correlated systems: Numerical
  methods}} (\bibinfo{publisher}{Springer}, \bibinfo{year}{2013}).

\bibitem{Feiguin2005}
\bibinfo{author}{Feiguin, A.} \& \bibinfo{author}{White, S.}
\newblock \bibinfo{title}{Time-step targeting methods for real-time dynamics
  using the density matrix renormalization group}.
\newblock \emph{\bibinfo{journal}{Physical Review B}}
  \textbf{\bibinfo{volume}{72}}, \bibinfo{pages}{020404(R)(1--4)}
  (\bibinfo{year}{2005}).

\bibitem{Nocera2016}
\bibinfo{author}{Nocera, A.} \& \bibinfo{author}{Alvarez, G.}
\newblock \bibinfo{title}{Symmetry-conserving purification of quantum states
  within the density matrix renormalization group}.
\newblock \emph{\bibinfo{journal}{Physical Review B}}
  \textbf{\bibinfo{volume}{93}}, \bibinfo{pages}{045137(1--11)}
  (\bibinfo{year}{2016}).

\end{thebibliography}

\newpage
\section*{Acknowledgements}
Work at Texas A\&M University (W. J. G. and M. C. A) was supported by NSF-DMR-1310008.  Work at Brookhaven National Laboratory (I. A. Z and A. M. T) was supported under the auspices of the US Department of Energy, Office of Basic Energy Sciences, under contract DE-SC00112704.  Work at Northeastern University (A. E. F.) was supported by the US Department of Energy, Office of Science, Basic Energy Sciences grant number DE-SC0014407.  Work at the ISIS neutron source was partly supported by the Science and Technology Facilities Council, STFC.  Access to DCS was provided by the Center for High Resolution Neutron Scattering, a partnership between the National Institute of Standards and Technology and the National Science Foundation under Agreement No. DMR-1508249.  We thank S. E. Nagler and R. Vega-Morales for helpful discussions about this work.

\section*{Author Contributions}
This work was designed by W. J. G., L. S. W., I. A. Z., and M. C. A.  The sample was synthesized by L. S. W.  The neutron scattering measurements at the ISIS neutron source were carried out by W. J. G., I. A. Z., F. D., and M. C. A.  The neutron scattering measurements at the NIST Center for Neutron Research were carried out by L. S. W., I. A. Z., Y. Q., J. R. D. C., and M. C. A.  The specific heat and magnetization data were measured by M. S. K.  All of the data were analyzed by W. J. G. with input from I. A. Z.  The tDMRG calculations were performed by A. E. F.  The theoretical model was developed by A. M. T.  The manuscript was written by W. J. G. with input from all authors.

\section*{Data Availability}
The data that support the findings of this study are available from the corresponding authors upon reasonable request.  Original time-of-flight data is available for the experiments at the ISIS neutron source at

\noindent http://dx.doi.org/10.5286/ISIS.E.42580328.

\section*{Competing Interests}
The authors declare that they have no competing interests.

\renewcommand{\thetable}{S\arabic{table}} 
\setcounter{table}{0}
\renewcommand{\theequation}{S\arabic{equation}} 
\setcounter{equation}{0}

\newpage
\section*{Supplementary Information \\ \vspace{20pt} Spinon Confinement and a Sharp Longitudinal Mode in Yb$_2$Pt$_2$Pb in Magnetic Fields}

\subsection*{Supplementary Note 1}

The observation of a continuum of scattering for all fields $\mu_0 H<2.3~\rm T$ necessarily indicates the presence of multi spinon states.  The energy of any two spinon state is given by the sum of the two spinon energies while the wavevector of the two spinon state is given by the vector sum of the momenta of each.  The energy and momentum of the two spinon state must conserved with respect to the change in energy and momentum between the incident neutron's initial and final states.  The possible energy and momentum of each spinon is given by the single spinon dispersions
\begin{equation}\label{Disp}
E_{\mathrm{p, h}}\left(\mathbf{q}\right)=\pm\left(I^2\sin ^2 \left(\pi \mathbf{q}_L\right)+\Delta_\mathrm{S}^2\cos ^2 \left(\pi \mathbf{q}_L\right)\right)^{1/2}
\end{equation}
where $E_{\mathrm{p,h}}$ is the energy for a ``particle'' and ``hole,'' \textit{i.e.} a spin-up spinon and a spin-down spinon.  As discussed in the main text, $I=0.485~\mathrm{meV}$ and $\Delta_\mathrm{S}=0.095~\mathrm{meV}$ for \YPP.

The envelope of the continuum is determined by the extremal two spinon states.  In zero field, there are three extremal cases.  In one case, if the hole momentum $\mathbf{q}_\mathrm{h}=0$, all of the momentum  $\mathbf{q}_{\mathrm{2spin}}=\mathbf{q}_\mathrm{p} +\mathbf{q}_\mathrm{h}$ of the two spinon state will be carried by the particle, $\mathbf{q}_{\mathrm{2spin}}=\mathbf{q}_{\mathrm{p}}$.  The energy of the two spinon state $E_{\mathrm{2spin}}$ will given by $E_{\mathrm{2spin}}=E_{\mathrm{h}}\left(0\right)+E_{\mathrm{p}}\left(\mathbf{q}_{\mathrm{2spin}}\right)$.  This is degenerate with the case where $\mathbf{q}_{\mathrm{2spin}}=\mathbf{q}_\mathrm{h}$ and $\mathbf{q}_\mathrm{p}=0$.  The third case has the particle and hole sharing the momentum equally, $\mathbf{q}_\mathrm{p}=\mathbf{q}_\mathrm{h}=\mathbf{q}_{\mathrm{2spin}}/2$, so that the two spinon dispersion is given by $E_{\mathrm{2spin}}=E_{\mathrm{p}}\left(\mathbf{q}_{\mathrm{2spin}}/2\right)+E_{\mathrm{h}}\left(\mathbf{q}_{\mathrm{2spin}}/2\right)$.  All other two spinon states have energies that fall between these possibilities.  They are sketched in Fig.1(B,D) of the main text.

Application of a field along the $z$ direction large enough for the Zeeman energy $g\mu_B HS^z$ to close the gap $\Delta_\mathrm{S}$ in the single spinon dispersion changes the extremal combinations of two spinons.  There are far more possible two spinon states, since there are now more routes to creating particle-hole paris as the hole (or particle, depending on the sign convention) dispersion is emptied with increasing field.  The boundaries of the two spinon continuum that are sketched in Fig. 2(D-I)  of the main text are arrived at by fixing the particles (holes) at momentum \kf\ and $1-$\kf\ and creating excitations with holes (praticles) at \kf\ and $1-$\kf, fixing the particle to have 0 momentum, fixing the particle to have momentum $\mathbf{q}_\mathrm{p} = 1$ and allowing the particle and hole to share the momentum equally. 

\subsection*{Supplementary Note 2}

The spectra for momentum along the chain direction \ql\ in a magnetic field (main text, Fig. 2(D-F)) are presented in Fig.~\ref{Spinon}(A-D) without the continuum boundaries.  Data at $\mu_0 H=1.9~\rm T$ are also shown.  The spectrum at 1.9 T is consistent with 1.5 and 1.7 T -- all are in the same antiferromagnetic phase.  The 4.0 T data in Fig.~\ref{Spinon}(E) are the same as in zero field, but with a \ql\ dependence that reflects the polarization factor of the Ising magnetic moments in the scattering plane, perpendicular to the applied magnetic field.

\subsection*{Supplementary Note 3}

The color plot used in the antiferromagnetic phase diagram shown in Fig.~3 of the main text was obtained from the specific heat.  We present the same color plot here in Fig.~\ref{PD} without the phase lines determined from magnetization.  Measurements of the specific heat $C_{\mathrm{p}}$ were made at many fixed temperatures as a function of a magnetic field  parallel to the crystal (110) direction.  An example of such a measurement at $T=0.35~\rm K$ is shown in Fig. ~\ref{SH}.  To enhance contrast for the color plot, the quantity plotted in the phase digram is the field derivative of the specific heat, ${\rm d}C_{\mathrm{p}}/{\rm d}\mu_0H$.  An example of the derivative at $T=0.35~\rm K$ is shown in Fig.~\ref{SH}.  Many such measurements are combined to make Fig.~\ref{PD}, which compares well with the phase diagrams determined previously. ~\cite{Ochiai_JPSJ_2011, Shimura_JPSJ_2012, Kim_PRL_2013}

\subsection*{Supplementary Note 4}

The spectra for momentum along  \qh, perpendicular to the chain direction (main text, Fig. 3(B-D)) are shown in Fig.~\ref{Mode_noFit}(A-D) without the mode position or dispersion fits.  Data at $\mu_0 H=1.9~\rm T$ are also shown.  The spectrum at 1.9 T is consistent with 1.5 and 1.7 T and is not shown in the main text.  The 4.0 T data in Fig.~\ref{Spinon}(E) are the same as in zero field, but with a \qh\ dependence that reflects the polarization factor of the Ising magnetic moments in the scattering plane, perpendicular to the applied magnetic field.  In Fig.~\ref{Mode_WithFit}, we show the same spectra with the mode position and dispersion fit over the entire measured range of \qh.  The dispersion and fits shown in Fig.3(B-D) of the main text are the same, but are only displayed for $\mathbf{q}_{HH}>0$ in order to show both the mode in the color plot and  in the analyzed data simultaneously.  The fits themselves were performed over the entire  range of \qh.

\subsection*{Supplementary Note 5}

The scattering geometry of our measurements orients the (1{\=1}0) crystal direction vertically, parallel to the magnetic field, with the (110) direction in the horizontal scattering plane.  Measurements made in a field $\mu_0 H>2.3~\rm T$ will quench all excitations in the Yb sublattice with moments oriented along (1{\=1}0), while having no effect on the perpendicular sublattice. Using such a measurement as a background for lower fields isolates  the fluctuations parallel to (1{\=1}0).  Plotting the total scattered intensity as a function of \qh\ at \ql$=1$ (main text, Fig.~3(F)) demonstrates the wavevector dependence of the scattering as \bq\ is rotated from the $c$-axis, ($\mathbf{q}_{HH}=0$) towards the (110) direction ($\mathbf{q}_{HH} \gg \mathbf{q}_L$). We  find a constant scattered intensity, and since neutrons only scatter from components of the local magnetization density perpendicular to \bq,~\cite{Squires_book_1978} this demonstrates that the fluctuating magnetic moments always have the same projection on the scattering vector and are thus always perpendicular to \bq. The same data plotted for a fixed field of 4 T show that the scattering precisely follows the geometric projection of the moments in the scattering plane onto the scattering vector, $I\propto \mathbf{q}_L^2/(2\mathbf{q}_{H}^2+\mathbf{q}_L^2)$, providing further confirmation that the fluctuations are purely longitudinal.

\subsection*{Supplementary Note 6}

The field dependence of the magnetization matches the field dependence of 2\kf\ in 1D systems~\cite{Giamarchi_book_2004}.  In  Fig.~\ref{KF}, we show the correspondence between the magnetization measured at $T=0.150~\rm K$, below the spinon gap $\Delta_\mathrm{S}$ with 2\kf\ calculated from where the chemical potential $g\mu_B HS^z$  crosses the single spinon dispersions, given for particles and holes by Eqn.~\ref{Disp} with $I=0.485~{\rm meV}$ and $\Delta_\mathrm{S} = 0.095~{\rm meV}$.  The magnetization has several kinks corresponding to metastability in the underlying incommensurate antiferromagnetic order, demonstrated in Fig.4(E,F) of the main text.

\subsection*{Supplementary Note 7}

Our description of the longitudinal interchain mode requires fitting the strengths of the relevant interchain couplings, $\mathcal{J}^{\perp}_n$, with $n=1$ for nearest neighbor coupling, $n=2$ for next nearest neighbors, \textit{etc}~\cite{Schulz_PRL_1996}.  The fitting function must be normalized to the total interchain exchange couplings $\mathcal{J}^{\perp}_{\mathrm{tot}}=\sum_{n=1}^5\mathcal{J}^{\perp}_n$ and include a gap $\Delta_\mathrm{M}$.  We give the dispersion by $E_{\mathrm{mode}}\left(\mathbf{q}_{HH}\right)^2=\Delta_\mathrm{M}^2\left(1-\frac{1}{3\mathcal{J}^{\perp}_{\mathrm{tot}}} \sum_{n=1}^5\mathcal{J}^{\perp}_n \cos\left(2\pi n \mathbf{q}_{HH}\right)\right)$, which includes 5 terms in the sum to account for the periodicity of the underlying antiferromagnetic order.   Defining $\mathcal{J}^{\mathrm{rel}}_{n} = \mathcal{J}^{\perp}_n/\mathcal{J}^{\perp}_{\mathrm{tot}}$, we vary $\mathcal{J}^{\mathrm{rel}}_{n}$ in our fits.  The results of the fit, along with the standard deviations of the fits are given in Table~\ref{Jtable}.

\begin{table}[H]
\begin{tabular}{|c||c|c|c|c|c|c|}
\hline
$\mu_0 H~\left(\rm T \right)$ & $\Delta_\mathrm{M}~\left({\rm meV}\right)$ & $\mathcal{J}^{\rm rel}_{1}$ & $\mathcal{J}^{\rm rel}_{2}$ & $\mathcal{J}^{\rm rel}_{3}$ & $\mathcal{J}^{\rm rel}_{4}$ & $\mathcal{J}^{\rm rel}_{5}$\\

\hline
\hline
1.0  & $0.174 \pm 0.001$ & $1.96 \pm 0.09$ & $-1.96 \pm 0.09$ & $-0.882 \pm 0.07$& $-0.248 \pm 0.06$ & $-0.119 \pm 0.06$\\

\hline
1.5 & $0.167 \pm 0.0006$ & $0.171 \pm 0.03$ & $1.01 \pm 0.03$ & $-0.0433 \pm 0.03$ & $-0.263 \pm 0.03$ & $-0.0258 \pm 0.03$\\

\hline
1.7 & $0.182 \pm 0.0007$ & $-0.100 \pm 0.03$ & $0.826 \pm 0.03$ & $0.190 \pm 0.03$ & $-0.119 \pm 0.03$ & $-0.0734 \pm 0.03$\\

\hline
\end{tabular}
\caption{Relative strength of the interchain couplings giving the longitudinal interchain mode dispersion.}
\label{Jtable}
\end{table}

\newpage

\newpage

\newpage

\newpage
\newpage

\renewcommand{\thefigure}{S\arabic{figure}}
\setcounter{figure}{0}

\begin{figure}
\includegraphics[height=80mm]{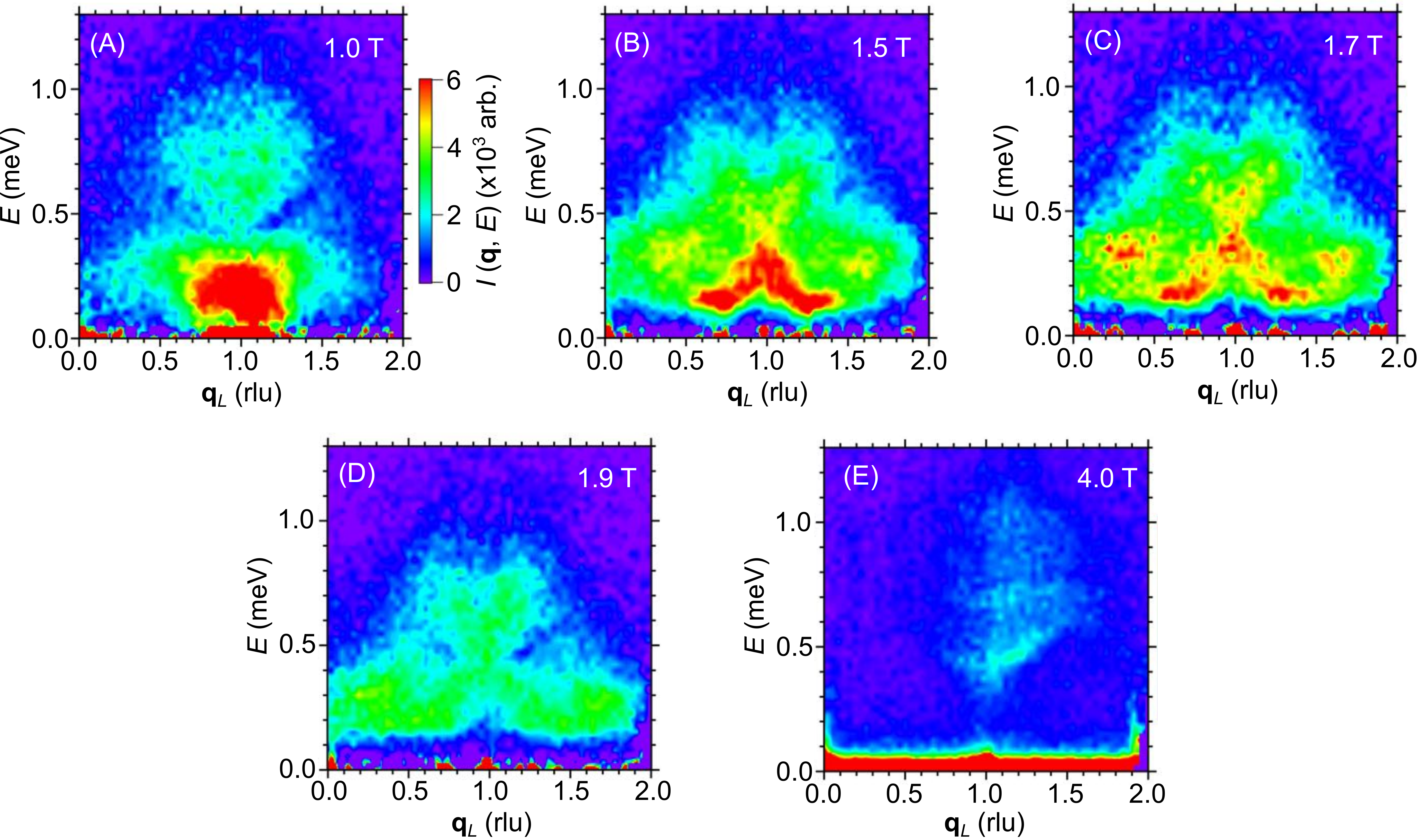}
\caption{\label{Spinon}\textbf{Spinons in \YPP\ in Magnetic Field}. The spectra for momentum along the chain direction \ql\ for fields of (A) $\mu_0 H=1.0~\rm T$, (B) 1.5 T, (C) 1.7 T, (D) 1.9 T, and (E) 4.0 T. Color scale is the same for all panels as part (A).  Data in panels A-D are the same as Fig. 2 of the main text, but without the continuum boundaries.}
\end{figure}

\begin{figure}
\includegraphics[width=80mm]{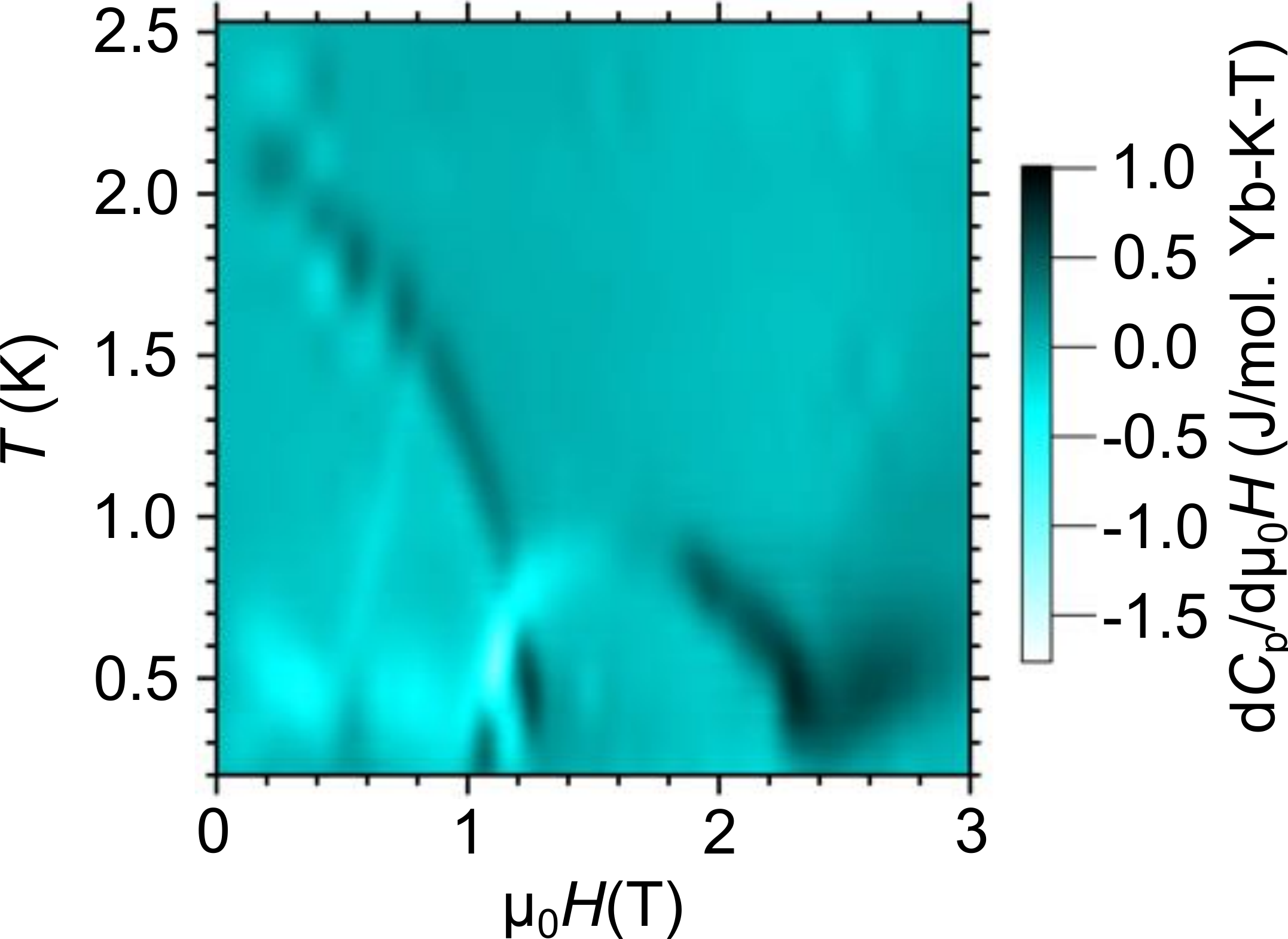}
\caption{\label{PD}\textbf{Antiferromagnetic Phase Diagram of \YPP}. The antiferromagnetic phase diagram of \YPP\ from measurements of the field derivative of the specific heat, an example of which is shown in Fig.~\ref{SH}.}
\end{figure}

\begin{figure}
\includegraphics[width=80mm]{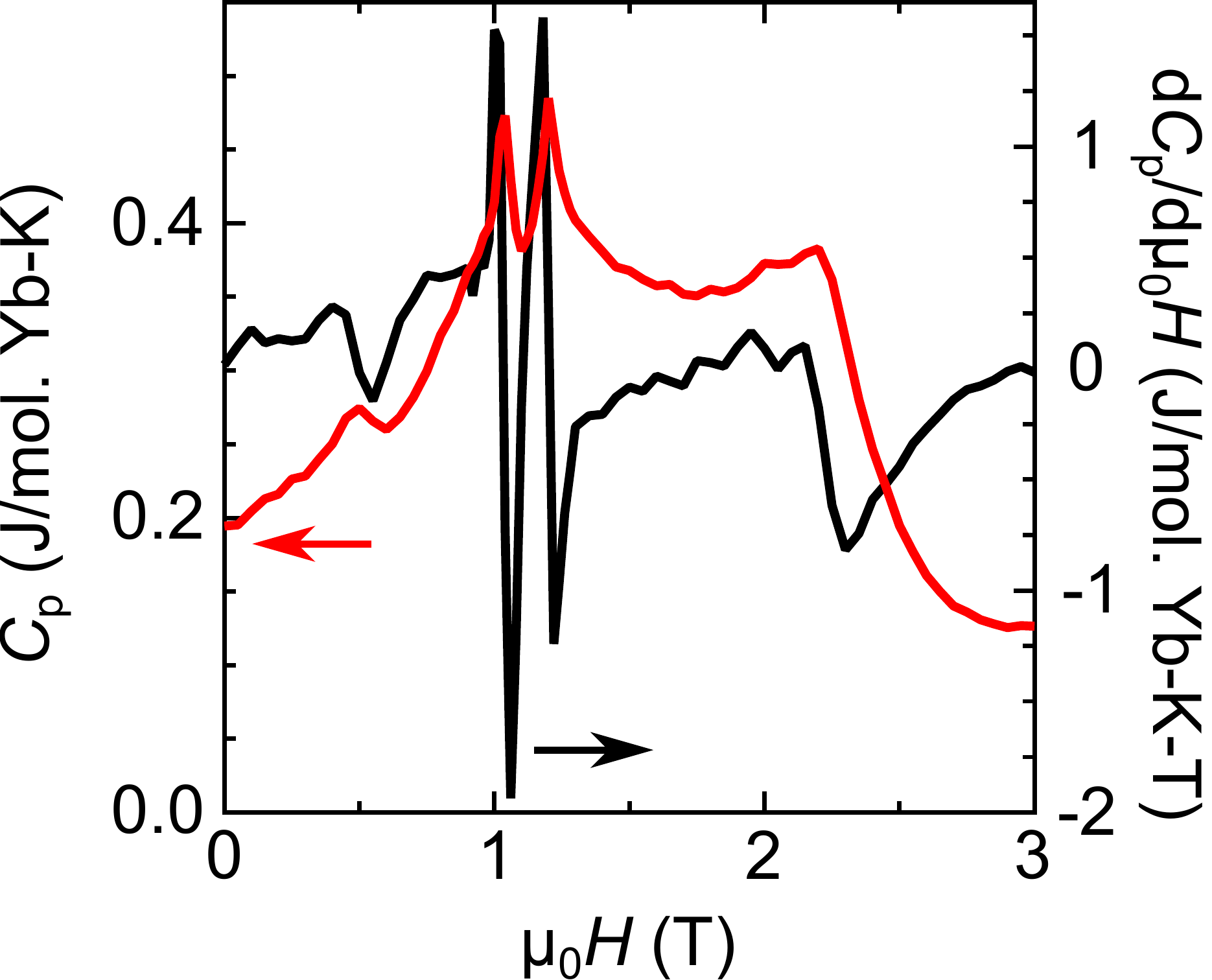}
\caption{\label{SH}\textbf{Specific Heat of \YPP}. The magnetic field dependence of the specific heat $C_\mathrm{p}$ of \YPP\ measured at $T=0.35~\rm K$ (red, left axis).  Also shown are the field derivative of the specific heat ${\rm d}C_\mathrm{p}/{\rm d}\mu_0 H$ (black, right axis) at the same temperature.}
\end{figure}

\begin{figure}
\includegraphics[height=80mm]{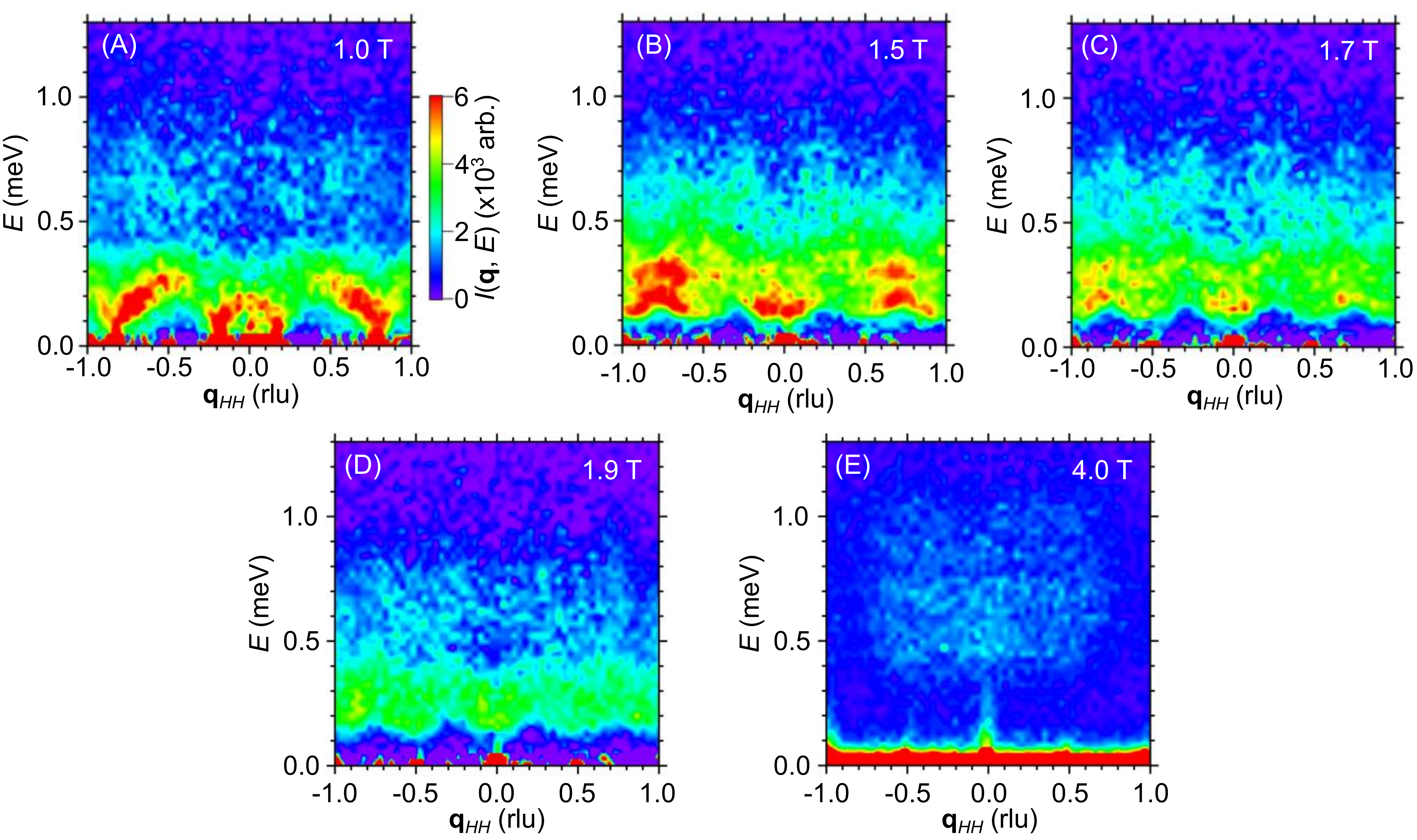}
\caption{\label{Mode_noFit}\textbf{Evolution of Interchain mode in \YPP }. The spectra for momentum perpendicular to the chain direction \qh\ for fields of (A) $\mu_0 H=1.0~\rm T$, (B) 1.5 T, (C) 1.7 T, (D) 1.9 T, and (E) 4.0 T. Color scale is the same for all panels as part (A).  Data in panels A-D are the same as Fig.~3 of the main text, but the mode position and dispersion fits are not shown.}
\end{figure}

\begin{figure}
\includegraphics[height=40mm]{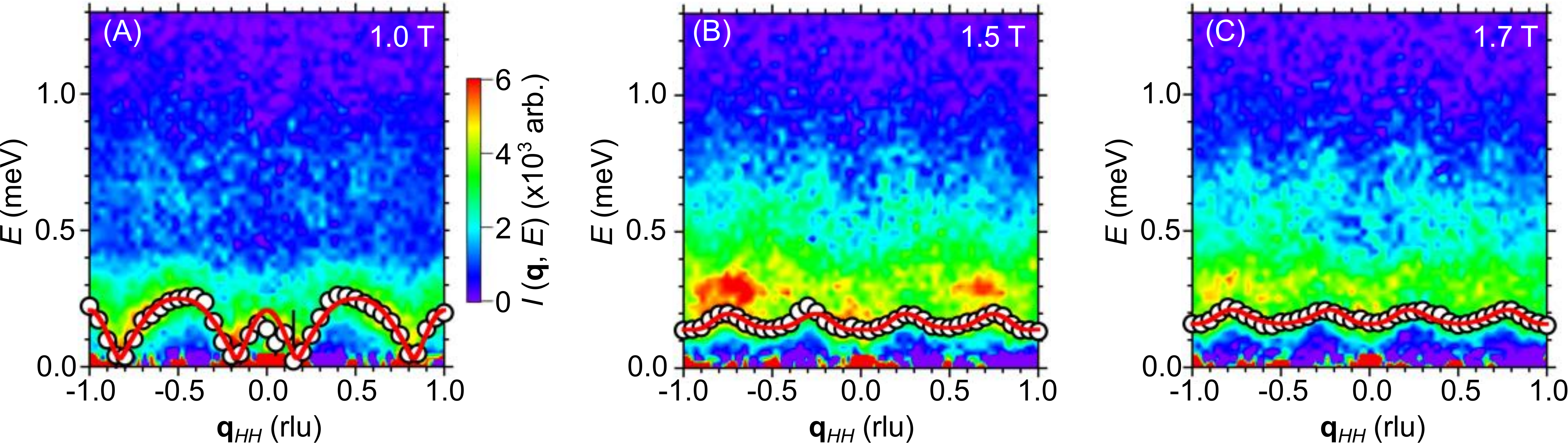}
\caption{\label{Mode_WithFit}\textbf{Evolution of Interchain mode in \YPP }. The spectra for momentum perpendicular to the chain direction \qh\ for fields of (A) $\mu_0 H=1.0~\rm T$, (B) 1.5 T, and (C) 1.7 T.  Color scale is the same for all panels as part (A).  Data are the same as Fig.~3 of the main text, but the mode position and dispersion fits are shown from $-1<\mathbf{q}_{HH}<1$ rlu.  Error bars represent one standard deviation.}
\end{figure}

\begin{figure}
\includegraphics[width=80mm]{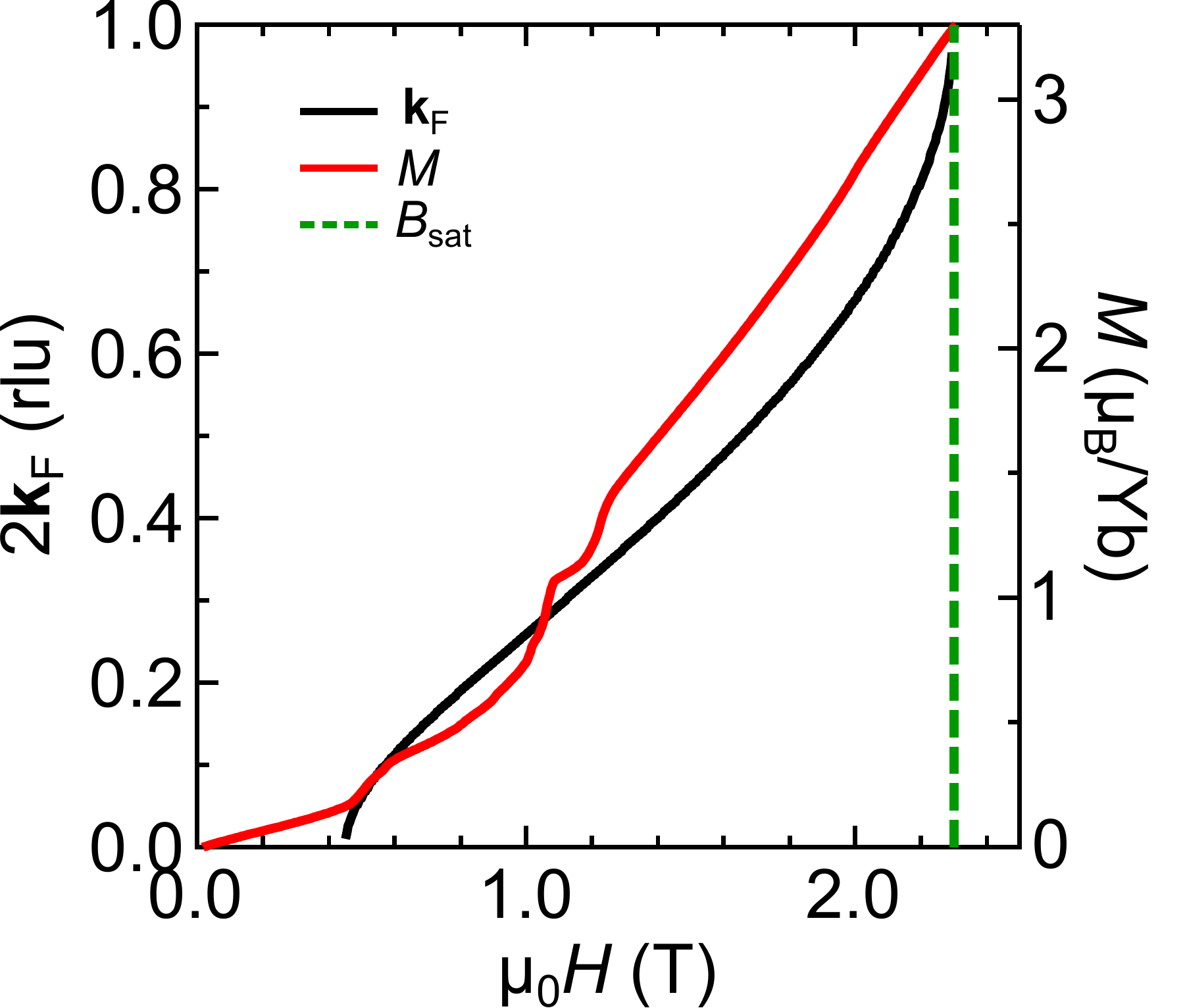}
\caption{\label{KF}\textbf{Fermi wavevector and Magnetization}. Two times the fermi wavevector \kf\ (black, left axis) and the magnetization $M$  (red, right axis) as a function of magnetic field along the (110) direction.  Magnetization was measured at $T=0.150~\rm K$.  The saturation field (green dashed line) is at $\mu_0 H_{\mathrm{sat}}=2.3~\rm T$.}
\end{figure}

\end{document}